\documentclass{emulateapj}
\usepackage[]{natbib}

\shorttitle{UV-optical colours of nearby early-type galaxies}
\shortauthors{S. Kaviraj et al.}

\begin{document}

\title{UV-optical colours as probes of early-type galaxy
evolution}

\author{S. Kaviraj,\altaffilmark{1} K. Schawinski,\altaffilmark{1} J. E.
G. Devriendt,\altaffilmark{1,9} S. Khochfar,\altaffilmark{1} S.-J.
Yoon,\altaffilmark{2,1} S. K. Yi,\altaffilmark{2,1} J.-M.
Deharveng,\altaffilmark{3} A. Boselli,\altaffilmark{3}
T. Barlow,\altaffilmark{4} T. Conrow,\altaffilmark{4} K.
Forster,\altaffilmark{4} P. Friedman,\altaffilmark{4} D. C.
Martin,\altaffilmark{4} P. Morrissey,\altaffilmark{4} S.
Neff,\altaffilmark{5} D. Schiminovich,\altaffilmark{6} M.
Seibert,\altaffilmark{4} T. Small,\altaffilmark{4}
T.Wyder,\altaffilmark{4}
L. Bianchi,\altaffilmark{7} J. Donas,\altaffilmark{3} T.
Heckman,\altaffilmark{7} Y.-W. Lee,\altaffilmark{2} B.
Madore,\altaffilmark{8} B. Milliard,\altaffilmark{3} R. M.
Rich\altaffilmark{2}\& A. Szalay\altaffilmark{7} }

\altaffiltext{1}{Department of Physics, University of Oxford,
Oxford OX1 3RH, UK} \altaffiltext{2}{Center for Space
Astrophysics, Yonsei University, Seoul 120-749, Korea}
\altaffiltext{3}{Laboratoire d'Astrophysique de Marseille, 13376
Marseille Cedex 12, France} \altaffiltext{4}{California Institute
of Technology, MC 405-47, Pasadena, CA 91125}
\altaffiltext{5}{Laboratory for Astronomy and Solar Physics, NASA
Goddard Space Flight Center, Greenbelt, MD 20771}
\altaffiltext{6}{Department of Astronomy, Columbia University, MC
5246, New York, NY 10027} \altaffiltext{7}{Department of Physics
and Astronomy, Johns Hopkins University, Baltimore, MD 21218}
\altaffiltext{8}{IPAC,770 S. Wilson Ave., Pasadena, CA 91125}
\altaffiltext{9}{Observatoire Astronomique de Lyon, 9 avenue
Charles André, 69561 Saint-Genis-Laval Cedex, France}


\begin{abstract}
We have studied $\sim$ 2100 early-type galaxies in the SDSS DR3
which have been detected by the GALEX Medium Imaging Survey (MIS),
in the redshift range $0<z<0.11$. Combining GALEX $UV$ photometry
with corollary optical data from the SDSS, we find that, at a 95
percent confidence level, \emph{at least} $\sim$ 30 percent of
galaxies in this sample have $UV$ to optical colours consistent
with \emph{some} recent star formation within the last Gyr. In
particular, galaxies with a $NUV-r$ colour less than 5.5 are
\emph{very} likely to have experienced such recent star formation,
taking into account the possibility of a contribution to $NUV$
flux from the UV upturn phenomenon. We find quantitative agreement
between the observations and the predictions of a semi-analytical
$\Lambda$CDM hierarchical merger model and deduce that early-type
galaxies in the redshift range $0<z<0.11$ have $\sim $ 1 to 3
percent of their stellar mass in stars less than 1 Gyr old. The
average age of this recently formed population is $\sim$ 300 to
500 Myrs. We also find that `monolithically' evolving galaxies,
where recent star formation can be driven \emph{solely} by
recycled gas from stellar mass loss, \emph{cannot} exhibit the
blue colours ($NUV-r<5.5$) seen in a significant fraction ($\sim$
30 percent) of our observed sample.
\end{abstract}




\section{Introduction}
One of the most important unresolved debates in contemporary
astrophysics concerns the formation mechanism of early-type
galaxies. In the classical `monolithic' model
\citep[e.g.][]{Larson74,Chiosi2002}, early-type stellar
populations form in a single, short, highly efficient burst of
star formation at high redshift $(z>>1)$, which is followed by
passive ageing to present day. Such a scenario can explain many of
the observed properties of early-type galaxies, both at local and
high redshift \citep{Chiosi2002}, without invoking hierarchical
merger-driven processes
\citep[e.g.][]{Toomre1972,Toomre77,KWG93,SP99,Benson2000,Cole2000,Hatton2003,Khochfar2003}.
Observational evidence that might \emph{conclusively} rule out
either the monolithic or the merger-based scenario has remained
elusive (see \citet{Peebles2002} for a critical review of the two
paradigms).

The properties of optical colour-magnitude relations (CMRs) of
early-type galaxies have often been taken as evidence in favour of
the monolithic scenario. The lack of redshift evolution of the
slope and scatter in optical CMRs
\cite[e.g.][]{BLE92,Bower98,Ellis97,Stanford98,Gladders98,VD2000}
is consistent with a high formation redshift ($z>2$), as is the
evolution in their zero-point. However, the predicted star
formation histories (SFHs) of early-type galaxies in the merger
paradigm make such a conclusion far less clear cut. For example,
the predicted SFHs of (cluster) early-types in the merger scenario
are \emph{quasi-monolithic}, with an \emph{overwhelming} majority
of the stellar mass forming before a redshift of 1
\citep{Kaviraj2005a}. A careful treatment in the merger framework
suggests that, the observed optical CMR can \emph{also} be
reconciled comfortably with merger models. From a photometric
point of view, optical colours are, at best, \emph{degenerate}
with respect to the competing theories of early-type galaxy
formation and it is very difficult to discriminate between them
using optical colours alone \citep{Kaviraj2005a}.

Evidence does exist for morphological evolution in galaxies,
suggesting that formation mechanisms of early-types are \emph{at
least not uniquely monolithic}. Although approximately 80 percent
of galaxies in the cores of present day clusters have early-type
morphologies \citep{Dressler80}, a higher fraction of spiral
galaxies have been reported in clusters at $0.3<z<0.8$
\citep[e.g.][]{BO84,Dressler97,Couch98,VD2000}, along with
increased rates of merger and interaction events
\citep[e.g.][]{Couch98,VD99}. This is supported by recent results
which suggest that the mass density on the red sequence (which is
dominated by early-type systems) has doubled since $z=1$
\citep{Bell2004}. The strengths of age-sensitive spectral indices
observed in some early-type systems require luminosity-weighted
ages which are consistent with the presence of at least some
recent star formation (RSF) in these systems
\citep[e.g.][]{CRC2003,Trager2000a,Trager2000b,Proctor2002}.
Furthermore, some nearby early-type galaxies, such as NGC 5128
\citep[e.g.][]{Rejkuba2001,Peng2003b,Yi2004,Kaviraj2005b,Rejkuba2004},
NGC 205 \citep{Hodge73,Burstein1988}, NGC 5102
\citep{Pritchet79,Burstein1988,Deharveng1997} and certain
early-type systems recently mapped by the SDSS
\citep{Fukugita2004} exhibit unambiguous signs of significant
ongoing or recent star formation. Others, such as NGC 2865
\citep{Bica1987}, NGC 5128 \citep{Schiminovich94}, NGC 3921
\citep{Schweizer1996} provide direct residual evidence of the
recent interactions that created them (see also papers by the
SAURON collaboration e.g. Falcon-Barroso et al. 2005 and
references therein). It is worth noting that, contrary to the
traditional notion of early-type galaxies being (cold) gas-poor
systems, observational evidence over the last thirty years (e.g.
from $IRAS$) have shown that this is \emph{not} the case. Cold
gas, which supplies the fuel for star formation, has been detected
in significant numbers of nearby early-type systems
\citep[e.g.][also see the review by
\citet{Knapp1999}]{Knapp1989,Knapp1996,Young2005}. Thus it is
perhaps correct to \emph{expect} RSF in early-type galaxies,
rather than consider them inert, passively evolving systems!

Given the apparently-evolving early-type fraction and accumulating
spectro-photometric evidence for RSF in nearby early-type
galaxies, it is vital to establish whether RSF may only be a
sporadic feature of a few early-types or a more widespread
phenomenon in the early-type population. The confirmation of
low-level star formation \emph{in a significantly large sample of
low-redshift early-type galaxies} would put powerful constraints
on models for their formation.

Spectroscopic indicators of RSF are already available, such as the
commonly used $H_{\beta}$ index, higher order Balmer lines such as
$H_{\gamma}$ and $H_{\delta}$ and the D4000 break. From a
photometric point of view, rest-frame ultraviolet flux is highly
sensitive to RSF. Given the apparent degeneracy in optical
colours, ultra-violet photometry offers the best route to finding
tell-tale signatures of RSF in early-type systems at low redshift
- if indeed such star formation is present in these systems!

In a recent work, \citet[][hereafter FS2000]{Ferreras2000} studied
a sample of early-type galaxies in the cluster Abell 851 at
$z=0.41$. They used $F300W$ (rest-frame near-ultraviolet ($NUV$),
2000 angstroms) and optical photometry to perform one of the first
studies of the $NUV$-optical CMR in a sample of early-type
galaxies. They found that the slope and scatter of the
$NUV$-optical CMR was consistent with some early-types having
$\sim$ 10 percent of their stellar mass in stars younger than
$\sim$ 500 Myrs. Detailed modelling of this data
\citep{Ferreras2002} indicated that secondary bursts of star
formation at low redshifts lead to a natural explanation of the
large scatter in the $NUV$-optical CMR.

More recently, \citet{Deharveng2002} studied the rest-frame flux
around $2000$ angstroms from a sample of 82 nearby early-types
using data from the \emph{FOCA}, \emph{SCAP} and \emph{FAUST}
experiments. The focus of this study was primarily to investigate
the far-ultraviolet ($FUV$) emission of early-type galaxies,
thought to originate from old extreme horizontal branch (EHB)
stars and their progeny \citep{Yi97}. However, their analysis
suggested that the $2000-V$ colours in some early-type galaxies
are significantly bluer than would be expected from old
populations alone and cannot be explained without invoking some
RSF.

Low redshift photometry in the far-ultraviolet ($FUV$; 1530
angstroms) and near-ultraviolet ($NUV$; 2310 angstroms) passbands
from the GALEX mission (Martin et al. 2005), unprecedented both in
terms of its quality and quantity, provides a unique opportunity
to study the RSF-sensitive $UV$ emission from a variety of nearby
early-type galaxies across a range of luminosities and
environments. In this paper we study $\sim$ 2100 early-type
galaxies detected by GALEX, in the redshift range $0<z<0.11$. We
focus mainly on the $NUV$ passband, because the $FUV$ is sensitive
not only to RSF but also to the $UV$ upturn flux from EHB stars
which could be expected in old early-type populations. The $NUV$
is less sensitive to $UV$ upturn and therefore a better RSF
indicator.

A preliminary study of the $UV$ emission of early-types, based on
GALEX detections of SDSS early-type galaxies listed in the catalog
of Bernardi et al. (2003d, B2003 hereafter) was presented in
\citet{Yi2005}. Comparing the photometry of early-types in this
sample to the spectral energy distribution (SED) of a strong
nearby $UV$-upturn galaxy (NGC 4552), they did not find more than
2 galaxies (out of 162) which fit the typical UV-optical shape of
a system with a significant amount of $UV$ upturn flux. They
concluded that the large scatter in the $NUV-r$ colours cannot be
generated from a scatter in late-stage stellar evolution alone. We
also direct readers to \citet{Boselli2005} who were the first to
perform a detailed study of the $UV$ CMR in the Virgo cluster.
However, we note that our sample covers a wide variety of
environments and we sample almost two magnitudes deeper (in
$r$-band) in dense environments \cite[see][]{Schawinski2005}
compared to the Virgo sample of \citet{Boselli2005}.

We begin this study by describing the construction of an
early-type catalog, similar to that used in \citet{Yi2005}, but in
which the morphological classification of objects and removal of
potentially $UV$-contaminating AGN are performed in a more robust
manner. Using a simple parametrisation of the SFH, we identify
galaxies which appear to have had \emph{some} star formation
within the last Gyr. We use a semi-analytical $\Lambda$CDM model,
calibrated to accurately reproduce the (cluster) optical CMR in
the redshift range $0<z<1.27$ \citep[see][]{Kaviraj2005a}, to
reproduce the observed $NUV$ CMR. Finally, we discuss the
comparative roles of quiescent and merger-driven processes in
driving the residual star formation found in some of our
early-type sample and investigate whether an \emph{extreme}
monolithic scenario is capable of reproducing the $NUV$ colours of
large ($>L_*)$, blue ($NUV-r<5.5$) early-type galaxies in our
sample.


\section{Sample selection}

\subsection{Constructing an early-type catalog from the SDSS DR3}
B2003 were the first to construct an SDSS catalogue of $\sim$ 9000
early-type galaxies, selected using a combination of SDSS pipeline
parameters (see also Bernardi et al. (2003b,c,d) for a
comprehensive study of the properties of galaxies in this
catalogue). Their catalogue contains galaxies with high $i$-band
concentration indices ($r_{50}/r_{90}> 2.5$), spectral
classifications typical of early-type galaxies, and where the
deVaucouleurs fit to the surface brightness profile is more likely
than the exponential fit. While such automated prescriptions are
very efficient at selecting a reasonably robust early-type galaxy
sample, it suffers from two shortcomings.

Firstly, it is apparent from visually inspecting a sample of
Bernardi early-types that it contains late-type contaminants, such
as sideways-on spirals, Sa-type galaxies and objects where a
central dominant bulge is surrounded by faint spiral features.
While clearly not early-type galaxies, these objects all pass the
selection criteria. The $UV$ emission from early-type galaxies is
predicted to be weak in any galaxy formation scenario. The
presence of late-type contaminants is therefore a particular
problem, since such galaxies will be substantially bluer than true
early-types and our conclusions regarding the presence of star
formation in the early-type population could be severely affected
\footnote{We must note, however, that given the size of the B2003
sample, performing a visual inspection of \emph{all} galaxies is
clearly impractical!}.

Secondly, \emph{spectral} selection criteria, such as those used
in B2003, immediately excludes any early-type galaxy which may
contain star formation and therefore spectral lines, such as those
reported by \citet{Fukugita2004}. Since one of our main aims is to
test galaxy formation models and attempt to discriminate between
the monolithic and merger driven paradigms, we must take special
care not to bias our sample towards one scenario at the outset!
Clearly, removing early-type systems with signs of star formation
will bias the sample towards the passive monolithic models and
make any conclusions regarding the recent star formation activity
in early-types less meaningful.

Since robust morphological classification inevitably requires
visual confirmation of early-type candidates, our approach to
constructing an early-type catalog relies more on visual
inspection of each candidate than automatic extraction based on
pipeline parameters. We make our initial selection of objects
based on a single SDSS parameter - $fracDev$, which is the weight
of the deVaucouleur's fit in the best composite (deVaucouleur's +
exponential) fit to the galaxy's image in a particular band. We
extract all galaxies which have $fracDev
> 0.95$ in $g,r$ and $i$ bands. The short wavelength $g$-band
traces disk or spiral arm structures, while $r$ and $i$ trace the
central bulge. The robustness of this procedure has been checked
by applying it to a sample of 200 randomly selected nearby
early-type galaxies from the SDSS DR3. We find that 90 percent of
the early-types in this sample (confirmed by visual inspection)
have $fracDev>0.95$. This implies that, on average, this criterion
would pick up 90 percent of the galaxies in a typical sample of
SDSS galaxies. Reducing the value of $fracDev = 0.95$ would admit
more early-types but also increase contamination, resulting in a
larger sample of galaxies to be processed visually. A $fracDev$
value of 0.95 gives the best trade-off between selecting a sample
dominated by early-type objects and a sample with a significant
amount of contamination. Note that we have chosen not to make a
distinction between elliptical and S0 galaxies in this study since
it is difficult to discriminate between them effectively using the
SDSS images. It is worth noting, however, that these two
populations could behave differently \citep[e.g][]{Boselli2005}.

Having extracted our initial working sample of galaxies, we
perform a visual inspection of each early-type candidate. However,
our sample of galaxies spans a large range in luminosity and
redshift. Our ability to accurately classify any galaxy depends on
the clarity of its features on the SDSS image - this in turn
depends on its redshift and apparent luminosity. Thus we must
first determine the \emph{limiting} redshift and luminosity at
which we can \emph{trust} our classifications.

The average SDSS exposure time is 50 seconds. To perform this
check we use 24,000 second exposure COMBO-17 images of SDSS
galaxies, at redshifts between 0.11 and 0.13 (kindly provided by
Chris Wolf). The high exposure time of the COMBO-17 images allows
us to resolve all features within the galaxies in question. We
find that, for an apparent \emph{r}-band magnitude of 16.8, our
classification based on the SDSS images matches that based on the
COMBO-17 images. We therefore apply a redshift cut of 0.11 and an
apparent \emph{r}-band magnitude cut of 16.8 to our sample, to
extract a magnitude limited sample of $\sim$ 3500 SDSS early-type
galaxies (see \citet{Schawinski2005} for more details).


\subsection{Cross-matching with GALEX MIS detections}
\begin{figure}
\begin{center}
\includegraphics[width=3.5in]{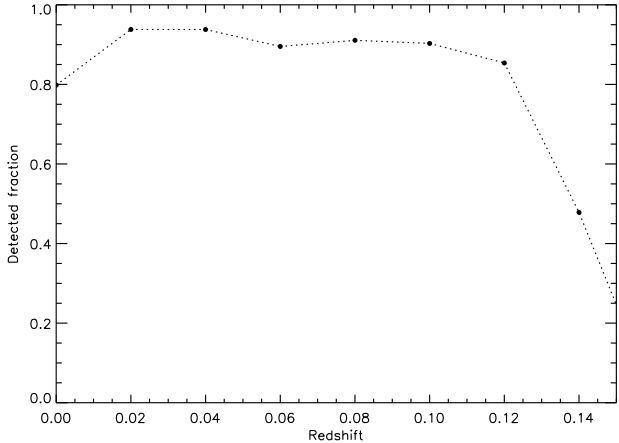}
\caption{\small{Fraction of SDSS early-types in MIS fields which
are detected by GALEX.}} \label{fig:mis_detection_rates}
\end{center}
\end{figure}

We direct readers to \citet{Martin2005} and \citet{Morrissey2005}
for a description of the scope and performance of the GALEX
satellite. The magnitude-limited SDSS early-type sample described
in the previous section is now cross-matched with data from 595
GALEX fields, imaged in the medium depth (MIS) mode. GALEX has a
fiducial angular resolution of 6 arcseconds. The positional
matching is performed within a more conservative 4 arcseconds. We
eliminate any GALEX objects that have multiple SDSS matches within
6 arcseconds because it is impossible to tell which companion
dominates the $UV$ flux detected by GALEX. We find that the
fraction of galaxies in our magnitude limited sample ($r<16.8$,
$z<0.11$) detected by GALEX is around 90 to 95 percent (Figure
\ref{fig:mis_detection_rates}). The detection rate drops sharply
after $z=0.11$.

\subsection{AGN diagnostics: optical analysis}
Since we are studying the (generally weak) rest-frame $UV$ flux
from early-type galaxies here, it is important to ascertain the
level of possible contamination of the $UV$ continuum due to
non-thermal emission from AGN, which are common in early-type
galaxies. \citet{Rich2005}, who employed a similar method to B2003
to extract a sample of early-type galaxies from the SDSS and study
their $UV$ emission detected by GALEX, were the first to perform
this additional check for AGN contamination, using emission lines
measured by the SDSS. Following their study, we perform a similar
analysis to gauge the contribution from AGN to the rest-frame $UV$
in our galaxies, and eventually remove those which exhibit signs
of hosting a strong AGN.

Type I (i.e. unobscured) AGN are largely removed by using the SDSS
spectral classification algorithm (by setting the QSO flag) during
our initial extraction of galaxies - any remaining Type Is are
then removed completely through the optical and radio analyses we
describe below. Type II (i.e. partly obscured) AGN can be
distinguished from normal (star-forming) galaxies using the
intensity ratios of pairs of strong emission lines \citep[][BPT
hereafter]{Baldwin1981}. Using the emission line ratios
[OIII/H$\beta$] and [NII/H$\alpha$] \citet[][K2003
hereafter]{Kauffmann2003} have used a BPT type analysis to
classify a large sample of SDSS galaxies into star-forming and
Type II AGN (seyferts, LINERs and transition objects). We identify
and remove potential AGN from our sample using the criteria
derived by K2003.

Before calculating line ratios, we must correct for possible
absorption in these line regions and extract the true emission
strengths of each line index. We first construct a library of
10,000 star formation histories (SFHs) in which each SFH is
modelled by an instantaneous starburst at $z=3$ followed by a
second instantaneous starburst, which is allowed to vary in age
and mass fraction. The choice of this model library is motivated
by the fact that our galaxy sample contains \emph{only} early-type
objects and we find that such model SFHs give excellent fits to
the stellar continua of galaxies in our sample. While studies such
as K2003 have to deal with the full spectrum of morphological
types, and thus require more elaborate model libraries, we tailor
our model library to the narrow set of morphologies spanned by our
sample.

The procedure for correcting for absorption proceeds in the
standard way, by first comparing the spectrum of each observed
galaxy (excluding major line regions) to the model library and
finding the best-fit model. The absorption strengths of the OIII,
H$\beta$, NII and H$\alpha$ lines in the best-fit model are then
computed and subtracted from the measured line strengths in the
observed spectrum. For each galaxy which exhibits all four
emission lines with a signal-to-noise ratio (S/N) greater than 3,
we construct the line ratios, place the galaxy on a plot of
[OIII/H$\beta$] vs. [NII/H$\alpha$] and classify it as either
star-forming or a Type II AGN, based on the criteria derived by
K2003.

\begin{figure}
\begin{center}
\includegraphics[width=3.5in]{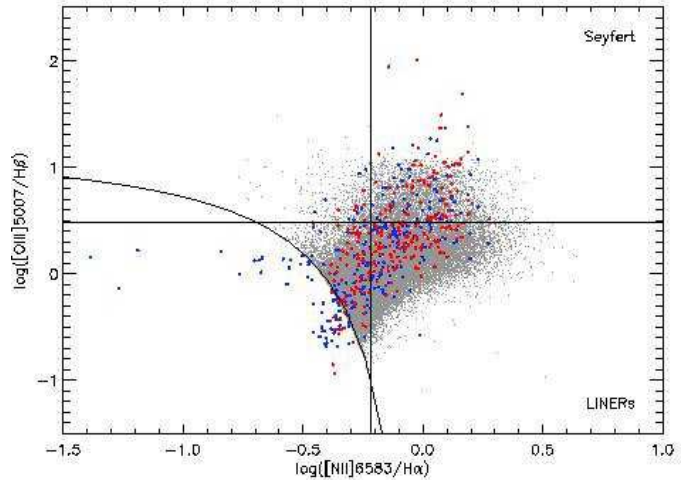}
\caption{Plot of [OIII/H$\beta$] vs. [NII/H$\alpha$] used to find
and remove AGN in our sample. The curved line represents the
demarcation between star-forming systems and Type II AGN -
galaxies are considered star-forming if they lie below this line.
Red points represent galaxies identified as early-type after
visual inspection. Blue points are galaxies identified as
contaminants in our initial sample of galaxies. The small grey
dots are K2003's SDSS Type II AGN catalog. It is worth noting that
the star-forming locus (see Figure 1 in K2003 for comparison) is
almost completely empty, which is expected since our sample is
composed exclusively of early-type galaxies. In addition, the
majority of galaxies which lie on the star-forming locus are
contaminants.} \label{fig:agn_bptplot}
\end{center}
\end{figure}

In Figure \ref{fig:agn_bptplot} we plot [OIII/H$\beta$] vs.
[NII/H$\alpha$] used to find and remove AGN in our sample. The
curved line represents the demarcation between star-forming
systems and Type II AGN - galaxies are considered star-forming if
they lie below this line. Red points represent galaxies identified
as early-type after visual inspection. Blue points are galaxies
identified as contaminants in our initial sample of galaxies. The
small grey dots are K2003's SDSS Type II AGN catalog. It is worth
noting that the star-forming locus (see Figure 1 in K2003 for
comparison) is almost completely empty, which is expected since
our sample is composed exclusively of early-type galaxies. In
addition, the majority of galaxies which do lie on the
star-forming locus are contaminants. This optical analysis
identifies $\sim$ 25 percent of our early-type sample as Type II
AGN.

\subsection{AGN diagnostics: radio analysis}
Galaxies which either do not have the four required emission
lines, or in which emission line detections have $S/N < 3$, cannot
be treated using the optical analysis discussed above. In most
studies these galaxies are assumed to have no contribution from
AGN. However, we opt not to make this assumption and confirm the
results of the optical analysis for galaxies which remain
unclassified using their radio emission. There are two radio
surveys which overlap with our early-type sample - the FIRST
survey \citep{Becker1995}, which has an angular resolution of
$\sim$ 5 arcseconds and a completeness limit of 1 mJy, and the
NVSS survey, which has an angular resolution of 45 arsceconds and
a completeness limit of 2.5 mJy.

In this study we use the FIRST survey due to its superior
resolution and depth - we find that including NVSS in our analysis
does not add to or alter our conclusions. The high resolution of
FIRST poses a slight problem for sources where the radio emission
is not very centralised. For example, galaxies with radio lobes
have multiple, spatially resolved detections in FIRST. We
therefore cross-match our early-type sample with FIRST detections
using a large search radius of 30 arcseconds. If there are
multiple detections within this search radius we add the fluxes of
all the detections corresponding to the early-type object in
question.

\begin{figure}
\begin{center}
\includegraphics[width=3.5in]{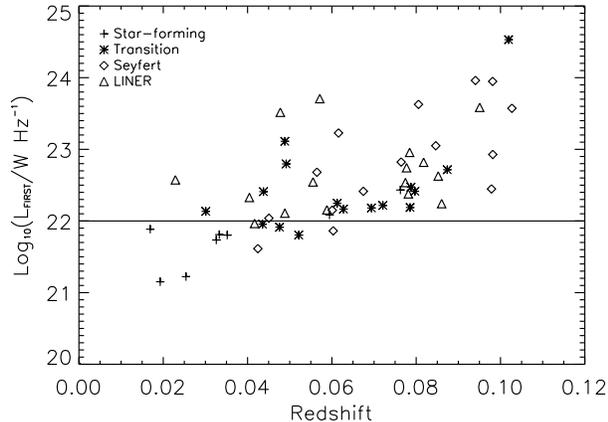}
\caption{Comparison of radio luminosities of galaxies classified
by the optical analysis as normal (crosses) and Type II AGN (all
other symbols, see legend for specific AGN classifications.) Type
II AGN tend to dominate this plot above a radio luminosity of
$\sim 10^{22} W Hz^{-1}$, while most of the normal (i.e.
star-forming) objects lie below this threshold.}
\label{fig:agn_radio}
\end{center}
\end{figure}

To identify a luminosity threshold above which radio activity
could be attributed to AGN, we compare the radio luminosities of
galaxies classified by the previous optical analysis as normal
(i.e. star-forming) to the luminosities of those classified as
Type II AGN. We show this comparison in Figure
\ref{fig:agn_radio}. Crosse show normal galaxies while other
symbols correspond to various classes of Type II AGN. We find that
Type II AGN tend to dominate this plot above a radio luminosity of
$\sim 10^{22} W Hz^{-1}$, while most of the normal (i.e.
star-forming) objects lie below this threshold. We use this
threshold and remove any galaxy which remains unclassified in the
optical analysis but shows a radio luminosity above $10^{22} W
Hz^{-1}$. This removes $\sim$ 3 percent of the sample of galaxies
which survive the optical analysis.

We must note, however, that galaxies with a high radio luminosity
do not necessarily host AGN. There is a degeneracy between radio
emission from AGN and from genuine star formation which is not
quantified accurately by using a single threshold value. By
applying this cut we might indeed exclude interesting early-types
galaxies which are genuinely star-forming. However, since the
radio analysis only affects a small fraction of our sample, we
\emph{include} it in our sample selection, as it would only make
our sample slightly more conservative in terms of star-forming
early-type galaxies.


\subsection{The final catalog}
Table \ref{tab:catalog_criteria} summarises all the criteria used
in the construction of our final catalog of early-type galaxies -
the final sample contains 2116 galaxies.

\begin{deluxetable}{cl}

\tablecaption{Summary of criteria used in constructing the
SDSS-GALEX early-type catalog.\label{tab:catalog_criteria}}

\tabletypesize{\small} \tablehead{\colhead{Criterion} &
\colhead{Reason}} \startdata

$r < 16.8$  & Robust morphology\\
            & (from COMBO-17 comparison)\\
$z < 0.11$  & Robust morphology,\\
            & detection rate stable at 90 percent\\
Emission line analysis & AGN removal for emission line\\
            & galaxies with $S/N > 3$\\
L(radio) $>10^{22} W Hz^{-1}$ & Further removal of possible AGN\\
            & not classified by the emission\\
            & line analysis \\

\enddata
\end{deluxetable}

\begin{figure}
\begin{center}
\includegraphics[width=3.7in]{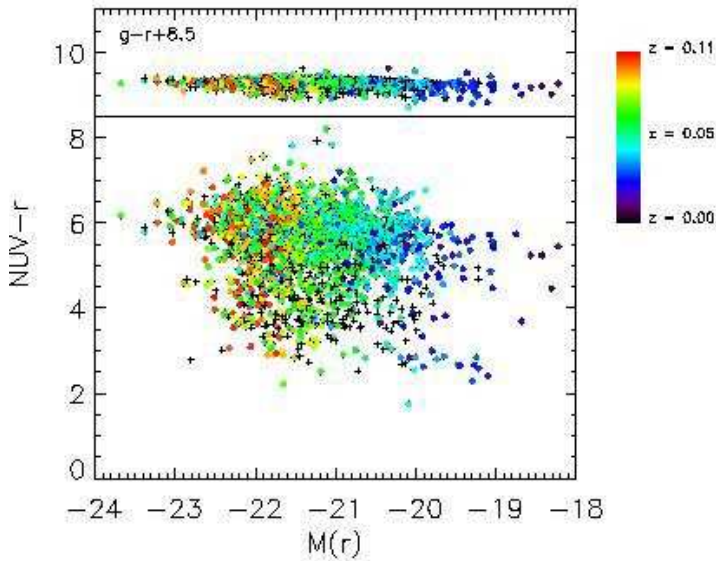}
\caption{The optical (top panel) and the $NUV-r$ (bottom panel)
colour-magnitude relations (CMRs) of our early-type galaxies. We
intentionally show the optical CMR on the same scale as its $UV$
counterpart, to highlight the significant difference in their
respective scatters. Galaxies are colour-coded according to their
redshifts. The small grey crosses represent galaxies which are
rejected based on the optical or radio analyses.}
\label{fig:opt_nuv_cmr}
\end{center}
\end{figure}

Figure \ref{fig:opt_nuv_cmr} shows the optical (top panel) and the
$NUV-r$ (bottom panel) colour-magnitude relations (CMRs) of our
early-type galaxies. Galaxies are colour-coded according to their
redshifts. The small grey crosses represent galaxies which are
rejected based on the optical or radio analyses described in the
previous sections. At first glance, the tight optical CMR typical
of early-type populations is in stark contrast to the the large
spread in colours evident in the $NUV$ CMR. While the scatter in
the optical CMR is approximately 0.05 mag, the spread in the $NUV$
colours is almost 6 mags with a (bi-weight) scatter of
approximately 1 mag. The properties of the $NUV$ CMR immediately
lead to two conclusions.

Firstly, it seems inconsistent with a `monolithic' evolution
scenario for all galaxies in our sample. If all early-types were
dustless simple stellar populations (as is frequently assumed from
their tight optical CMRs in the context of the monolithic
picture), we might expect the $NUV$ CMR to be \emph{just as tight}
as its optical counterparts. In reality, any age dispersion,
coupled with the presence of varying amounts of $UV$ upturn, might
be expected to induce some scatter in the $NUV$ CMR. A lower limit
to the $NUV-r$ colour in such a hypothetical monolithic CMR can be
estimated by looking at the $NUV-r$ colour of the nearby strong
$UV$-upturn galaxy, NGC 4552. We use a composite spectrum of NGC
4552, in which the $UV$ spectral ranges are constructed from $IUE$
and $HUT$ data (see caption of Figure 1 in \citet{Yi1998} for more
details on sources in the construction of this composite SED), to
estimate this lower limit. This composite SED yields a $NUV-r$
colour of 5.4. We should note, however, that the $UV$ component,
derived from $IUE$ and $HUT$ data, samples the \emph{$UV$ bright
core} of NGC 4552, so the value of 5.4 is likely to be
\emph{bluer} than the $NUV-r$ colour averaged over the entire
galaxy. This makes $NUV-r=5.4$ an optimistic estimate for the blue
end of a monolithic CMR. However, in a monolithic scenario all the
galaxies would have $NUV-r \geq 5.4$, which is clearly not the
case for the galaxies in our sample. We can therefore assume that
the excess scatter in the $NUV$ CMR is due, at least in part, to
some recent star formation.

Secondly, it is apparent that the \emph{red envelope} is not very
well-defined i.e. we do not see a sharp red edge to the $NUV$ CMR
as is frequently seen in optical CMRs. This is likely to be caused
by varying levels of dust (and therefore gas) in these galaxies
which \emph{smear out} the tightness in the red sequence.



Note that $NUV$ colours are a factor of 4 more sensitive to dust
than optical colours such as $(g-r)$, so the presence of a small
amount of dust can have a proportionately larger effect in the
$NUV$ than in the optical spectral ranges.

\begin{figure}
$\begin{array}{c}
\includegraphics[width=3.5in]{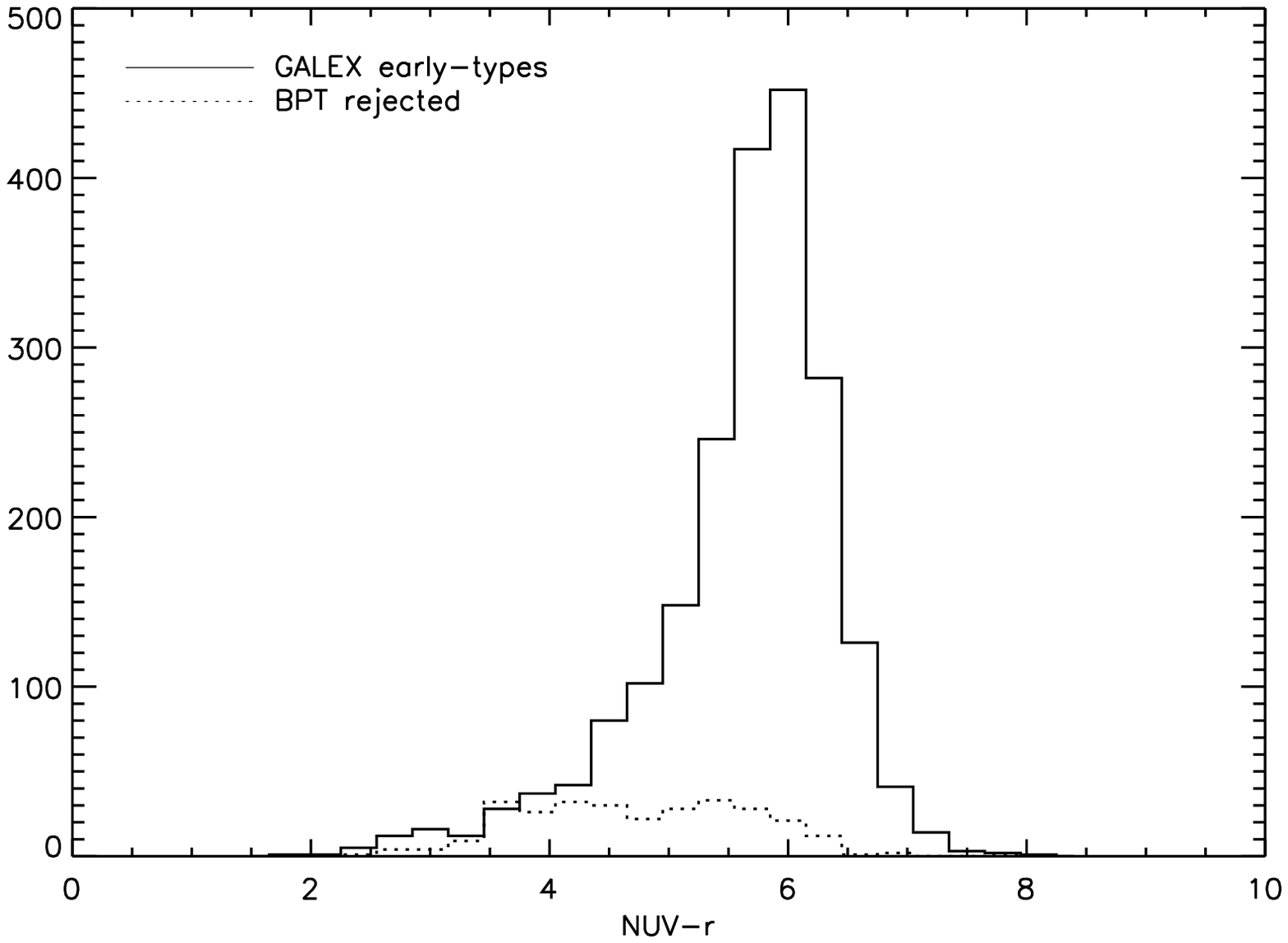}\\
\includegraphics[width=3.5in]{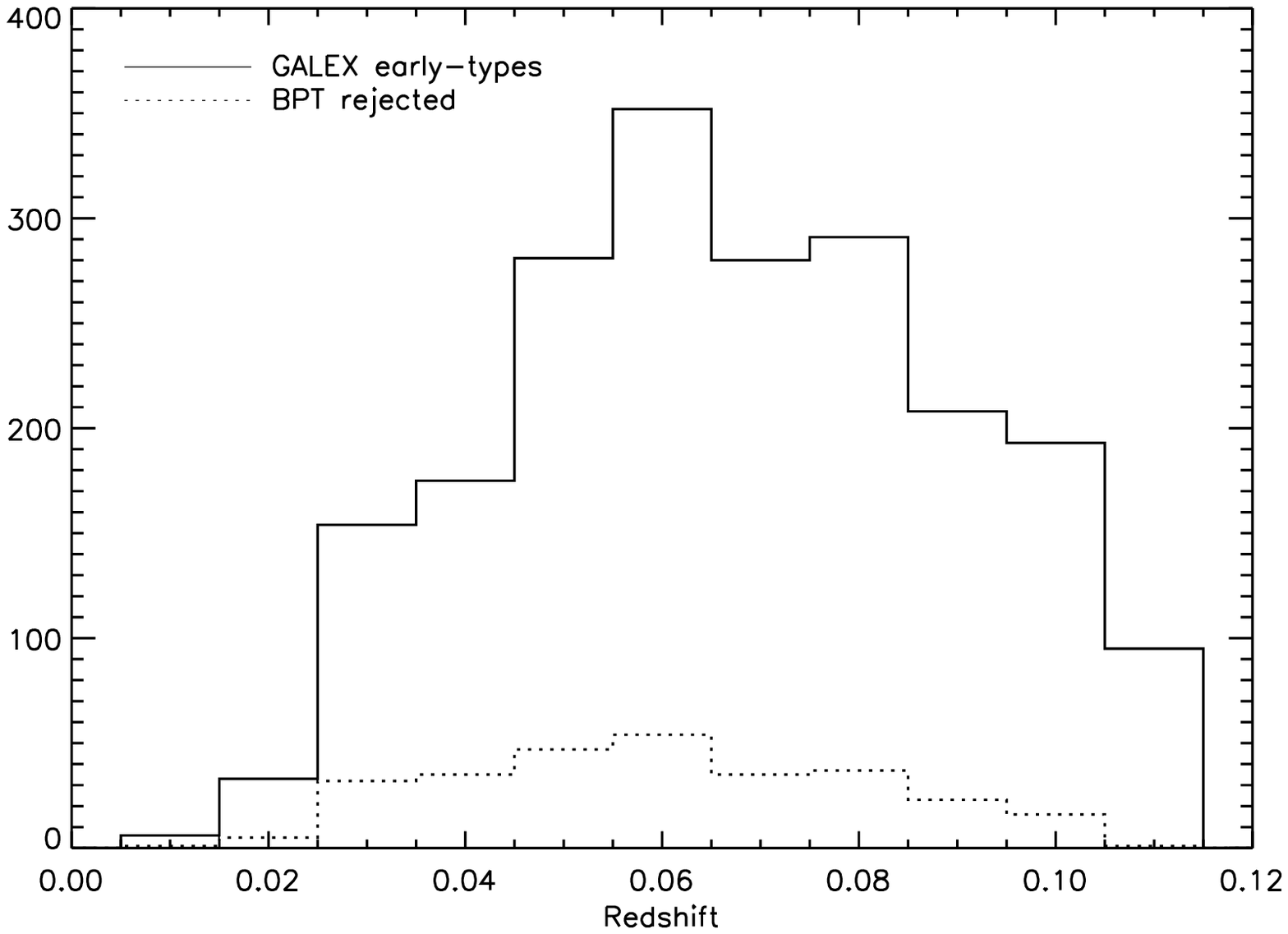}
\end{array}$
\caption{Comparison of BPT rejected galaxies to galaxies in our
final catalog. The BPT rejected galaxies prefer bluer $NUV-r$
colours, which is not unexpected since the AGN contribution to the
continuum would be expected to make the $NUV$ colour bluer. There
appears to be no bias with redshift - we do not expect the
fraction of AGN hosting galaxies to change significantly over this
redshift range.} \label{fig:bpt_rejected}
\end{figure}

\begin{figure}
\begin{center}
\includegraphics[width=3.5in]{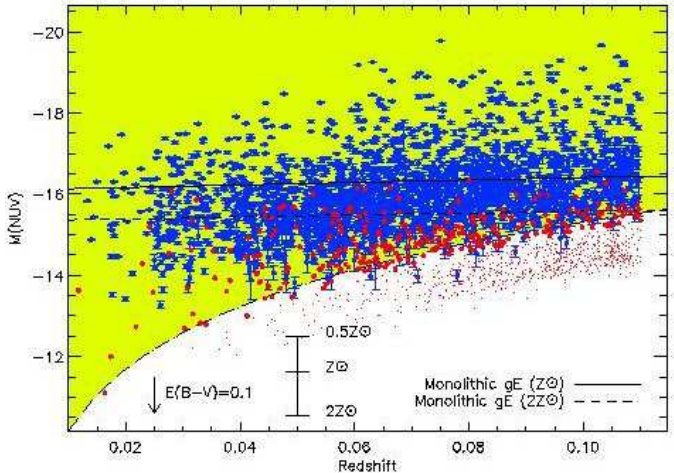}
\caption{SDSS early-types detected by GALEX: the absolute $NUV$
magnitude has to be in the yellow region for the galaxy to be
detected by GALEX. The detection limit corresponding to the MIS
fields is shown by the dashed curve. The absolute $NUV$ magnitude
of the detected early-types and their errors are shown by the blue
points. Also shown is the expected evolution of the absolute $NUV$
magnitude of a monolithic giant early-type galaxy ($M_{V}=-23$)
which formed at z=3 and has solar (solid line) and twice solar
(dashed line) metallicity.} \label{fig:detected_early_types}
\end{center}
\end{figure}

In Figure \ref{fig:bpt_rejected} we compare the BPT rejected
galaxies to galaxies in our final catalog. The BPT rejected
galaxies prefer bluer $NUV-r$ colours, which is not unexpected
since the continuum AGN contribution makes the $NUV$ colour bluer.
There appears to be no bias with redshift which fits our
expectation that the fraction of AGN hosting galaxies will not
change significantly over such a small redshift range. In Figure
\ref{fig:detected_early_types} we show the $NUV$ luminosities of
early-types in our sample as a function of redshift. The yellow
region shows the detectable $M_{NUV}$ space, assuming the fiducial
detection limit of the MIS fields ($m(NUV)=23$, dashed line). The
absolute $NUV$ magnitude of the detected early-types and their
errors are shown by the blue points. Also shown is the expected
absolute $NUV$ magnitude of a passively evolving giant early-type
galaxy ($M_{V}=-23$) which forms at $z=3$, assuming solar (solid
line) and twice solar (dashed line) metallicities. Since the
monolithic evolution shown is for an extremely bright early-type
galaxy, detections above this line immediately imply that, for at
least some early-types, a monolithic SFH is not consistent with
their detected $NUV$ flux.

As shown in Figure \ref{fig:mis_detection_rates}, approximately 10
percent of the SDSS early-types in the GALEX field-of-view remain
undetected. To explore the properties of these undetected
galaxies, we construct, for each non-detection, a \emph{synthetic}
absolute $NUV$ magnitude, based on the assumption that the galaxy
is dustless, has solar metallicity and is formed at $z=3$. We plot
these synthetic magnitudes as red points in Figure
\ref{fig:detected_early_types}. The filled red circles correspond
to undetected galaxies with $r < 16.8$ and the small red points
are non-detections with $r>16.8$. We find that, while some
non-detections are close to the detection limit, a significant
number of these galaxies \emph{should} be detected, under the
assumptions used to construct their synthetic magnitude. However,
both a super-solar metallicity and/or a small dust content can be
sufficient to push these galaxies below the detection limit - we
indicate this, in Figure \ref{fig:detected_early_types}, by
showing the predicted deviation of the $NUV$ luminosity from the
synthetic magnitude assumptions for a small dust content, e.g.
$E(B-V)=0.1$, and non solar metallicities.

\section{Quantifying the RSF in the GALEX early-types}
The aim of this section is to \emph{quantify} the RSF in our
early-type sample. In particular, we would like to establish how
many of our galaxies have photometry consistent with any star
formation within the last Gyr, at the 95 percent confidence level.

\begin{figure}
$\begin{array}{c}
\includegraphics[width=3.5in]{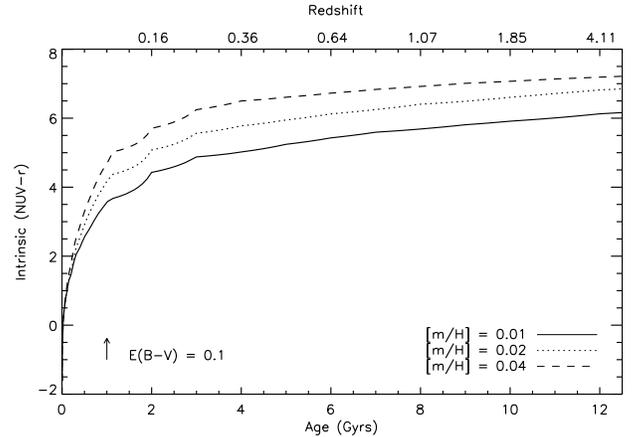}\\
\includegraphics[width=3.5in]{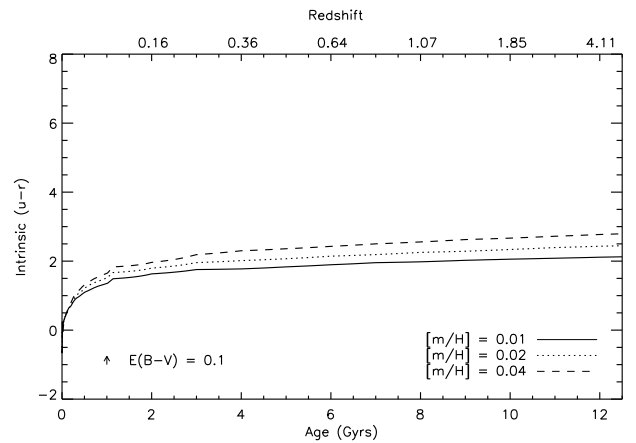}
\end{array}$
\caption{$NUV-r$ (top) and $u-r$ (bottom) colours of a
  simple stellar population (SSP) as a
  function of age. The solid line shows SSPs with half solar
  metallicity, the dotted line represents solar metallicity and the
  dashed line indicates twice solar metallicity. The evolution in the
  $NUV-r$ colour slows after 2 Gyrs because the $NUV$ flux
  from a recent starburst decays significantly in this timescale.}
\label{fig:ssp_nuv_evolution}
\end{figure}

Before we begin, it is instructive to consider the scope of the
$NUV$ in computing age determinations. Figure
\ref{fig:ssp_nuv_evolution} shows the intrinsic $NUV-r$ colour of
a simple stellar population (SSP) as a function of age and
metallicity (solar: dotted line, half-solar: solid line and
twice-solar: dashed line). The evolution in the $NUV-r$ colour
slows considerably after 1-2 Gyrs, as the $NUV$ flux from a recent
starburst decays significantly in this timescale - the $NUV$ is
therefore an effective age indicator for only $\sim$ 1 Gyr after
the starburst in question. In other words, we can expect the $UV$
(+ optical) photometry of our galaxies to be able to distinguish
only between galaxies which have had a very recent starburst,
within the last 1 Gyr and galaxies that havent. The $NUV$,
however, is blind to any star formation which is much older than 1
Gyr old.

We fit the observed colours of the GALEX early-types across the
$UV$ and optical spectral ranges to colours predicted by a library
of model SFHs, in which an initial starburst at high redshift
($z=3$) is followed by a second starburst which is allowed to vary
in mass fraction and age between $z=3$ and present day. Both
starbursts are assumed to be instantaneous. Each model in the
library has two main parameters: $t_{YC}$, the age of the second
starburst, in the range $10^{-2}$ to $15$ Gyrs and $f_{YC}$, the
mass fraction of stars formed in the second starburst, in the
range $10^{-4}$ to $1$ (`YC' stands for Young Component). We
explore models with metallicities $Z/Z_{\odot}$ in the range $0.5$
to $2.5$ and intrinsic $E(B-V)$ values in the range $0$ to $0.15$.
We use the standard dust prescription given in
\citet{Calzetti2000} to compute the reddening for a given value of
$E(B-V)$. The stellar models used to compute the model library in
this study \citep{Yi97,Yi2003} adopt the universal Salpeter IMF.
For galaxies without $FUV$ detection, we use a four colour
($NUV-r$, $g-r$, $r-i$ and $r-z$) comparison, while for those with
$FUV$ detection we use five colours ($FUV-r$, $NUV-r$, $g-r$,
$r-i$ and $r-z$). We perform a $\chi^2$ test with the appropriate
degrees of freedom by minimising the sum of the normalised
residuals

\begin{equation}
\chi^2 = \sum_n^{4/5}{\Big(\frac{C^{mod}_n-C^{obs}_n}{\sigma_n}\Big)^2},
\end{equation}

where $C^{mod}_n$ and $C^{obs}_n$ are the $n^{th}$ model and
observed colours respectively and $\sigma_n$ is the uncertainty in
the residual $C^{mod}_n-C^{obs}_n$. The average errors in the SDSS
optical magnitudes are approximately 0.01 mag. The average
uncertainty in the GALEX $FUV$ and $NUV$ magnitudes are $\sim
0.25$ mag and $\sim 0.15$ mag respectively. The uncertainties in
stellar models are taken to be approximately 0.05 mag for optical
passbands and 0.1 mag for the UV passbands \citep[see][]{Yi2003}.
Since, our observed sample contains galaxies at various redshifts,
models are redshifted to $z=0$ for each comparison, thus avoiding
uncertainties due to K-corrections. The derived model fits are
therefore in the \emph{rest-frame} of each observed early-type
galaxy.

\begin{figure}
\begin{center}
\includegraphics[width=0.5\textwidth]{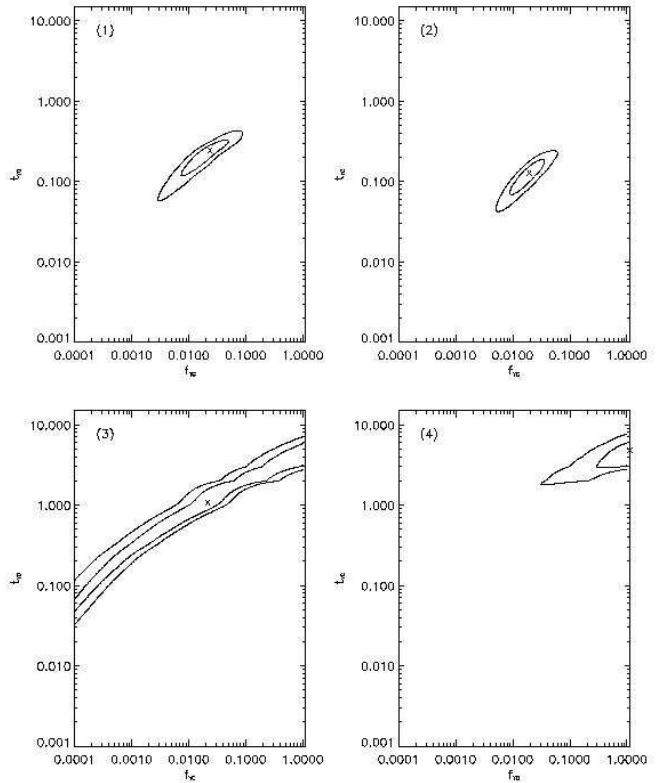}
\caption{Four examples of $\chi^2$ contour maps. The $\chi^2$
minimum is marked with `\textbf{x}' and we indicate the 68 percent
$(1 \sigma$) and 95 percent confidence contours. Galaxies 1 and 2
(top row) have clear signatures of RSF, since the 95 percent
contours are contained completely within $t_{YC} < 1
\textnormal{Gyr}$. Galaxy 3 shows a large age-mass degeneracy
because the $NUV$ flux from a small young starburst can be Galaxy
4 is clearly likely to have old or intermediate age populations
since its best-fit $t_{YC}$ is greater than 3 Gyrs and its 95
percent confidence contour does not extend far below 2 Gyrs.}
\label{fig:contour_map_examples}
\end{center}
\end{figure}

We use this procedure to derive best-fit values of
$(t_{YC},f_{YC})$ and $\chi^2$ likelihoods in $(t_{YC},f_{YC})$
space for each galaxy. Our entire sample can be broadly divided
into three categories - we illustrate this by showing the $\chi^2$
contour maps of four example galaxies in Figure
\ref{fig:contour_map_examples}. The $\chi^2$ minimum is marked
with \textbf{x} and we indicate the 68 percent and 95 percent
confidence contours. Galaxies 1 and 2 (top row) have clear
signatures of RSF, since their 95 percent contours are contained
completely within $t_{YC}<$ 1 Gyr. Galaxy 3 shows a large age-mass
degeneracy which appears in most cases because the $NUV$ flux from
a small young starburst can be indistinguishable from that due to
a larger but older starburst. Degenerate contours can also be the
result of large errors in the observed photometry which allow many
models to fit the data with values of (reduced) $\chi^2 < 2$.
Therefore, even though Galaxy 3 has a best-fit $t_{YC}$ around 1
Gyr, we cannot rule out the possibility of younger or older
populations being present, purely from its UV and optical
photometry. Finally, Galaxy 4 is likely to have old or
intermediate age populations, since its 95 percent confidence
contour lies exclusively at ages above 2 Gyrs.

We note that, although they used a similar model for their
analysis, the age-mass degeneracies in fits to our sample of
galaxies are smaller than those studied by FS2000. This is partly
due to the fact that optical uncertainties in SDSS measurements
are ten times smaller than those used by FS2000. Together with the
additional $FUV$ constraint (FS2000 only used rest-frame $NUV$),
this allows a much smaller part of the $(f_{YC},t_{YC})$ space to
fit the colours of the observed early-types.

\begin{figure}
\begin{center}
\includegraphics[width=3.5in]{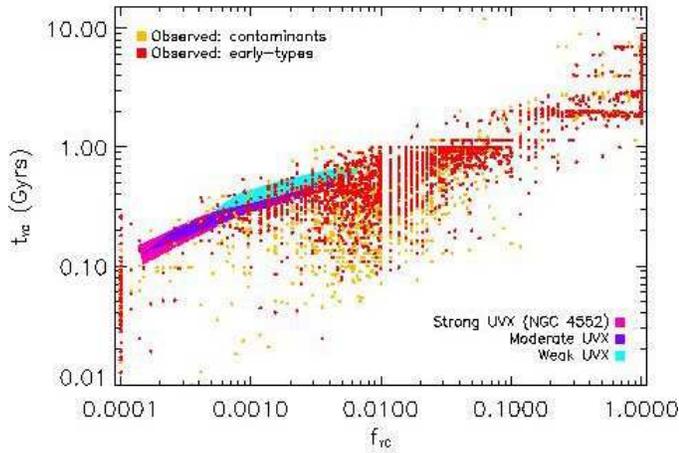}
\caption{\small{Best-fit values of $f_{YC}$ and $t_{YC}$.}}
\label{fig:ts_bestvalues}
\end{center}
\end{figure}

\begin{figure}
\begin{center}
\includegraphics[width=3.5in]{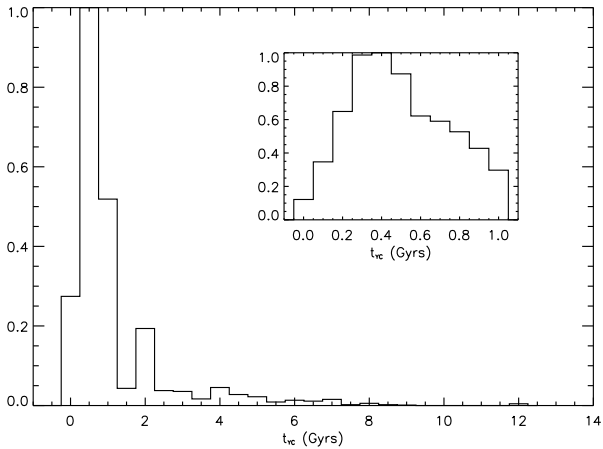}
\caption{Distribution of best-fit $t_{YC}$ values in our sample of
  galaxies - the inset shows the distribution of values restricted to
  the range $0<t^{best-fit}_{YC}<1$ Gyrs}.
\label{fig:tyc_histograms}
\end{center}
\end{figure}

In Figure \ref{fig:ts_bestvalues} we summarise the \emph{best-fit}
values of $(t_{YC},f_{YC})$ for our entire sample of galaxies. Red
points indicate the best-fit values for early-type galaxies in our
sample while orange points indicate best-fit positions of
contaminants (galaxies which are rejected by the visual inspection
or rejected by the AGN diagnostics). We find an increased scatter
of contaminants towards lower values of $t_{YC}$ (i.e. younger
ages). We also show the 95 percent confidence contours of a galaxy
with strong, moderate and weak UV upturn (UVX) flux. We generate
the strong UVX contour using the observed SED of NGC 4552. The
moderate and weak UVX contours are computed by reducing the UV
flux ($<3500$ angstroms) in the NGC 4552 SED by a factor of 2 and
6 respectively. In Figure \ref{fig:tyc_histograms} we show the
distribution of best-fit $t_{YC}$ values in our sample of galaxies
- the inset shows the distribution of values restricted to the
range $0<t^{best-fit}_{YC}<1$ Gyrs. Figures
\ref{fig:ts_bestvalues} and \ref{fig:tyc_histograms} suggest that
the UV + optical photometric properties of the \emph{majority} of
early-type galaxies in our sample are not compatible with a single
rapid burst at high redshift followed by passive evolution. Such a
conclusion is difficult to achieve with optical CMRs studied in
the past
\cite[e.g.][]{BLE92,Bower98,Ellis97,Stanford98,Gladders98,VD2000},
which are all consistent with purely passive monolithic scenarios.

We now classify our galaxies using their $\chi^2$ contours, as
follows: if the 95 percent confidence contour lies exclusively
below a $t_{YC}$ of 1 Gyr then we label the galaxy as \emph{young}
(e.g. galaxies 1 and 2 in Figure \ref{fig:contour_map_examples}).
If the 95 percent contour spans ages above and below 1 Gyrs, the
galaxy is classified as \emph{degenerate} (e.g. galaxy 3 in Figure
\ref{fig:contour_map_examples}). Finally, if the 95 percent
contour lies exclusively above an age of 1 Gyr then the galaxy is
classified as \emph{old} (e.g. galaxy 4 in Figure
\ref{fig:contour_map_examples}). As indicated in Figure
\ref{fig:ts_bestvalues}, some galaxies have best-fit values which
fall within the UVX contours defined by the observed SED of NGC
4552. These galaxies clearly have SEDs which resemble that of
typical UV upturn galaxies but they may be initially labelled as
\emph{young}. However, since it is impossible to determine whether
the shape of their SED is a result of RSF or UVX or a combination
of both, we \emph{re-label} these galaxies as degenerate.

\begin{figure}
\begin{center}
\includegraphics[width=3.5in]{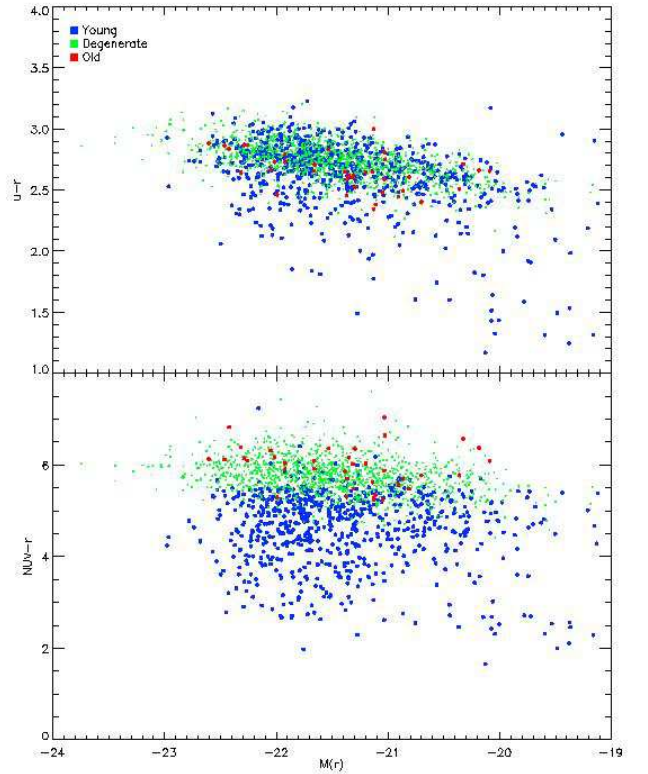}
\caption{The optical $u-r$ CMR (top panel) and the $NUV$ CMR
(bottom panel) for the GALEX early-types colour-coded using the
following classification scheme: if the 95 percent confidence
contour does not extend to ages higher than 2 Gyrs then the galaxy
is classified as \emph{young} (e.g. galaxies 1 and 2 in Figure
\ref{fig:contour_map_examples}); if the 95 percent confidence
contour spans ages below and above 2 Gyrs the galaxy is classified
as \emph{degenerate} (e.g. galaxy 3 in Figure
\ref{fig:contour_map_examples}); if 95 percent confidence contour
lies exclusively above an age of 2 Gyrs then the galaxy is
classified as \emph{old} (e.g. galaxy 4 in Figure
\ref{fig:contour_map_examples}). \emph{Young} galaxies are shown
in blue, \emph{intermediate} galaxies are shown in green and
\emph{old} galaxies are shown in red. Note that the photometry has
not been K-corrected because this requires the assumption of a
template - it is evident that an old passively evolving population
is not good template for the local early-type population, at least
for UV photometry.} \label{fig:cmr_comparison}
\end{center}
\end{figure}

In Figure \ref{fig:cmr_comparison}, we plot the $u-r$ and $NUV-r$
CMRs for the GALEX early-types and colour-code each galaxy using
this classification scheme. We find that the optical CMR is unable
to distinguish between different levels of RSF, since the various
classifications cannot be distinguished in the optical CM space.
However, the $NUV$ CMR shows a clear \emph{layered} structure. The
galaxies classified as \emph{young} (blue symbols) are bluest in
the $NUV-r$ colours and clearly separated from the \emph{old}
locus (red symbols) and \emph{degenerate} class of galaxies (green
symbols) forming an intermediate layer. Blue galaxies reside
almost exclusively below $NUV-r \sim 5.5$ (for comparison, the
$NUV-r$ colour of NGC 4552 is 5.4). Degenerate and old galaxies
lie almost exclusively above $NUV-r \sim 5.5$.

We note that this method indicates the presence of RSF in a very
general sense. Two instantaneous starbursts are clearly an
oversimplification of the true SFH of early-type galaxies. The
position of a galaxy in $(f_{YC},t_{YC})$ space is a function of
the \emph{shape} of its SED and our efforts in this section have
centred on approximating the shape of the true SED of each galaxy
(using a library of $f_{YC}$ and $t_{YC}$) through colour fitting.
Although the quantities $f_{YC}$ and $t_{YC}$ do not have a direct
physical meaning, the two-starburst method probes whether there
has been \emph{any} star formation in these systems within the
last Gyr. Since our classifications take account of the 95 percent
confidence contours, i.e. the part of parameter which enclose 95
percent of the probability, we are able to derive conclusions
about the presence of such RSF at the 95 percent confidence level.
Having identified the signature of RSF in an object, we must then
appeal to the various formation paradigms (see Sections 5 and 6
below) to motivate the star formation that generates the observed
RSF.


\section{The blue fraction}
\begin{figure}
\begin{center}
\includegraphics[width=3.5in]{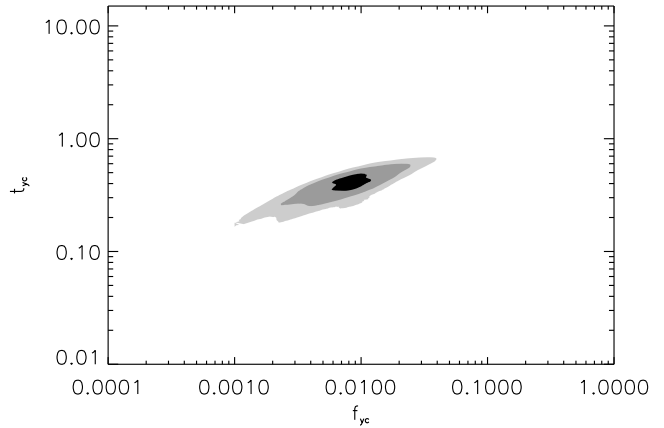}
\caption{Stacked likelihood map of blue galaxies, constructed by
adding the likelihood maps of each blue galaxy together. The black
region indicates the region where the average likelihood is higher
than 95 percent. The dark and light grey regions correspond to
regions where the probability is between 70 and 95 percent and 30
and 70 percent respectively. The black region represents the
values of $f_{YC}$ and $t_{YC}$ which are, on average, most likely
for the blue population.} \label{fig:lk_map_blue}
\end{center}
\end{figure}

Galaxies classified as \emph{young} and shown in blue in Figure
\ref{fig:cmr_comparison} are systems with a clear signature of
RSF. In Figure \ref{fig:lk_map_blue} we show a stacked likelihood
map of all \emph{blue} galaxies, constructed by adding the
likelihood maps of individual blue galaxies (c.f. top row of
Figure \ref{fig:contour_map_examples}) together. The black region
indicates the part of parameter space where the average likelihood
is higher than 95 percent. The dark and light grey regions
correspond to regions where the probability is between 70 and 95
percent and 30 and 70 percent respectively. Thus, the black region
represents the values of $f_{YC}$ and $t_{YC}$ which are, on
average, most likely for the blue population. The most likely
value of $f_{YC}$ is $\sim$ 1 percent and $t_{YC}$ is $\sim$ 0.5
Gyrs.

\begin{figure}
\begin{center}
\includegraphics[width=3.5in]{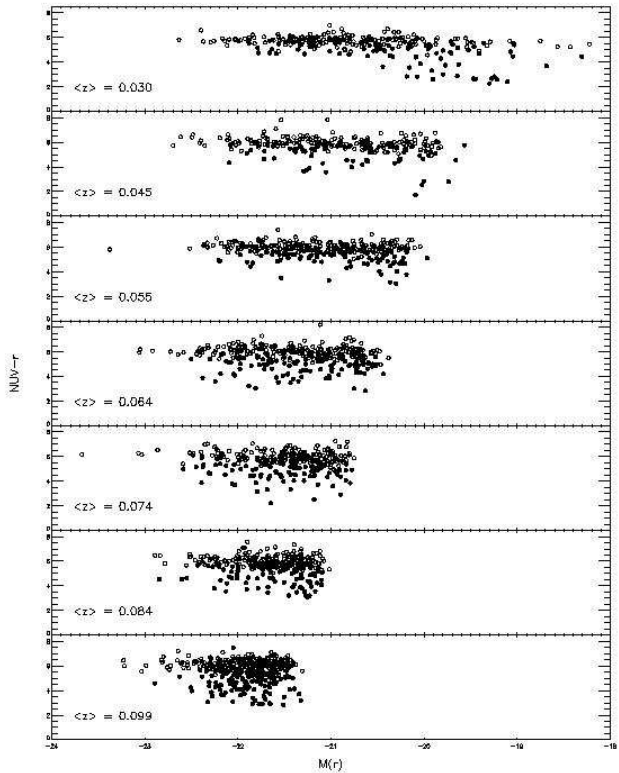}
\caption{The evolution of the NUV CMR with redshift. Galaxies
classified as \emph{young} in the colour-fitting analysis are
shown using filled circles.} \label{fig:blue_galaxies}
\end{center}
\end{figure}

In Figure \ref{fig:blue_galaxies}, we split the NUV CMR into
narrow redshift bins and indicate the blue population in each bin.
It is apparent from Figure \ref{fig:blue_galaxies} that, as a
result of constructing a (apparent) magnitude limited sample, we
are not sampling identical parts of the luminosity function at
different redshifts. Comparison of the blue fraction across our
redshift range is therefore meaningful only for the part of the
luminosity function which is sampled completely at all redshifts -
looking at the highest redshift bin this appears to be for
galaxies with $M(r) < -21.3$.

\begin{figure}
\begin{center}
\includegraphics[width=3.5in]{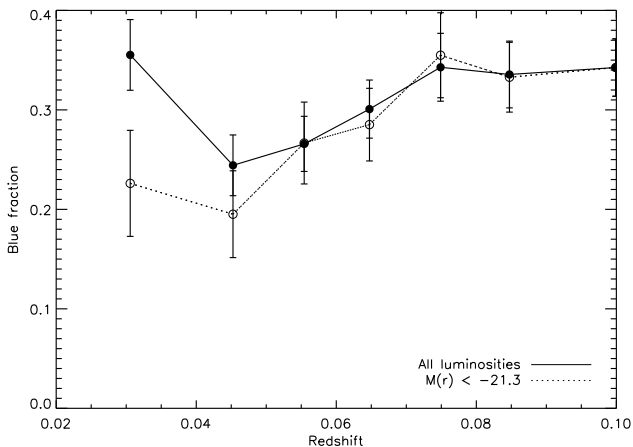}
\caption{Evolution of the blue fraction with redshift. The solid
curve takes into accounts all galaxies in each redshift bin (see
Figure \ref{fig:blue_galaxies}), while the dotted curve considers
only galaxies with $M(r) < -21.3$. Note that, due to arguments
given in Section 4, the blue fraction presented here is a
\emph{conservative estimate}.} \label{fig:blue_fraction}
\end{center}
\end{figure}

In Figure \ref{fig:blue_fraction}, we plot the evolution of the
blue fraction with redshift. Recall that $\sim 10$ percent of
early-type galaxies in the SDSS sample remain undetected by GALEX
(Figure \ref{fig:detected_early_types}). The true value of the
blue fraction should take these galaxies into account. We
therefore assume that none of the non-detections are blue and
adjust the blue fraction accordingly. This is a fairly strong
assumption because the undetectability of these galaxies is a
function of both their dust content and metallicity. It is
possible that some of these galaxies do have star formation but
are heavily dust-reddened and/or metal-rich. Nevertheless, without
their UV photometry this is a question we cannot answer.

The solid curve in Figure \ref{fig:blue_fraction} takes into
account all galaxies in each redshift bin, while the dotted curve
considers only galaxies with $M(r) < -21.3$. Note also that at
very low redshifts ($z<0.05$), SDSS has a \emph{bright} detection
limit - spectroscopy is not possible for bright galaxies ($M(r) <
-22.7$) which are extremely close ($z<0.05$). Hence we do not
properly sample the luminosity function in the first two redshift
bins, although the blue \emph{fraction} should remain unchanged
due to this undersampling. The red curve in Figure
\ref{fig:blue_fraction} indicates that the fraction of blue
systems for bright galaxies ($M(r) < -21.3$) varies between $\sim$
27 and 34 percent with an average value of $\sim$ 30 percent.
Within the errors there is negligible evolution in the blue
fraction within the redshift range $0<z<0.11$. Nevertheless, both
these points make our blue fraction a \emph{conservative
estimate.}

It is worth noting that the blue fraction computed in this study
is a \emph{lower limit} on the fraction of early-type galaxies
which contain RSF. It is possible that some of the galaxies
classified as \emph{degenerate} also contain a certain level of
RSF, but this is impossible to verify given their photometry
and/or observational uncertainties. In addition, we have assumed
that undetected ellipticals are not blue although it is clearly
possible that they \emph{have} some star formation but are heavily
dust-reddened or very metal-rich. This is, however, unquantifiable
and any correction to the blue fraction could be expected to be
negligible within errors.

\section{Comparison to the merger paradigm}
Since we have established that RSF \emph{does} exist in a
significant fraction of the early-type population, it is natural
to ask whether the properties of our local early-type sample can
be reproduced in the currently popular hierarchical merger
paradigm. We therefore explore the predicted $NUV$ CMR in the
semi-analytical framework using the GALICS model, which combines
large scale N-body simulations with analytical recipes for the
dynamical evolution of baryons within dark matter haloes. We
direct readers to \citet[][H2003 hereafter]{Hatton2003} for
specifics regarding the fiducial model. \cite{Kaviraj2005a} have
used this model to accurately reproduce the the observed optical
CMR in cluster early-type galaxies in the redshift range
$0<z<1.27$ (see their Section 2). We use the blind UV predictions
from this \emph{optically-calibrated model} to study the colours
of the GALEX early-type sample. Note that galaxy morphologies in
the model are defined by the ratio of $B$-band luminosities of the
disc and bulge components, which correlates well with Hubble type
\citep{Simien1986}. A morphology index is defined as

\begin{equation}
I = \exp{-L_B/L_D}
\end{equation}

so that pure discs have $I=0$ and pure bulges have $I=1$.
Following \citet{Baugh1996}, early-type galaxies have $I<0.507$
\footnote{In this scheme ellipticals have $I<0.219$, S0s have
$0.219<I<0.507$ and spirals have $I>0.507$.}.

\begin{figure}
\begin{center}
\includegraphics[width=3.5in]{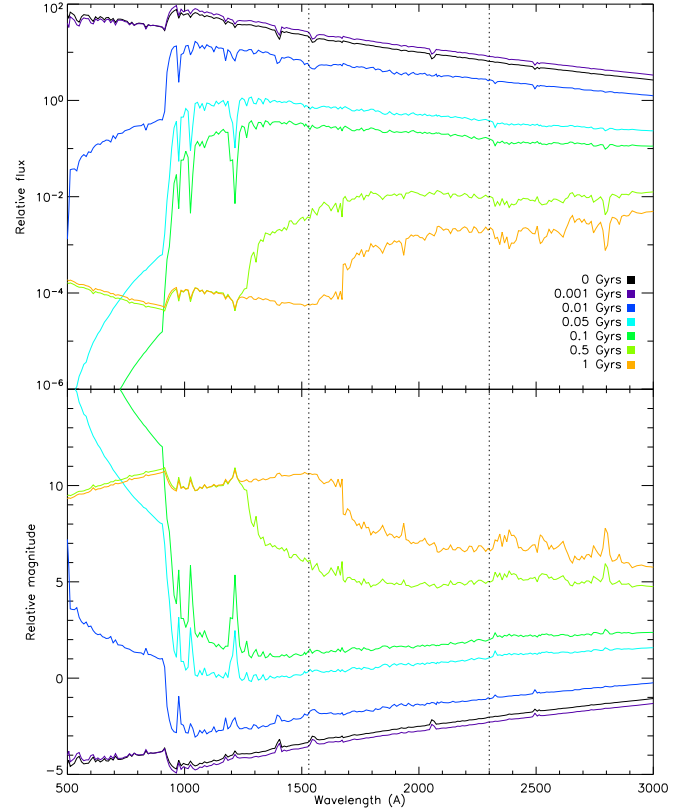}
\caption{Evolution of the UV flux with age. TOP: Relative UV
fluxes of simple stellar populations. BOTTOM: Relative magnitudes
of simple stellar populations shown in the top panel. The dotted
lines indicate the peak throughput of the $FUV$ and $NUV$ filters.
All models assume solar metallicities and are taken from the
public BC2003 distribution.} \label{fig:uv_flux_evolution}
\end{center}
\end{figure}

We approach the problem from the point of view of reproducing the
$(NUV-r)$ \emph{colour distribution} using an ensemble of model
galaxies in the redshift range spanned by the observed sample. In
traditional (optical) CMR studies, the colour distribution is
often parametrised by the slope and scatter of the correlation.
This is useful because the optical correlations are typically
\emph{tight} i.e. have small scatter. But since the $NUV$ CMR has
more structure than can be expressed through the slope and scatter
alone, matching only the slope and scatter within errors (which
are large due to the significant dispersion in colours) is no
longer sufficient.


\subsection{Treatment of very young stars}
Since we are modelling the $UV$ spectrum in this study, we must
address some additional issues regarding the flux from (very)
young stars in the model. The evolution of $UV$ flux with the age
of the $UV$-emitting young stars is highly non-linear. Figure
\ref{fig:uv_flux_evolution} shows the evolution of the $UV$
spectrum with age, on both flux and magnitude scales for a solar
metallicity stellar model from the public \citet{BC2003}
distribution. It is apparent that the $UV$ flux of a 1 Myr
population is $\sim$ 4-5 \emph{orders of magnitude} greater than
that of a 1 Gyr population. Like other models of galaxy formation,
star formation in GALICS is essentially proportional to the cold
gas available in the galaxy - stars are therefore produced
\emph{continuously}. In addition, the GALICS model provides high
age resolution (1 Myr) for very young stars. Therefore in a
significant number of galaxies, a residual fraction of very young
stars ($<10$ Myrs old) is present.

There are two main issues regarding such young stars. Very young
stars are typically embedded in birth clouds (BCs). During this
embedded phase the attenuation due to dust in the ambient ISM is
compounded by the dust present in the outer HI envelopes of the
BCs. Thus for embedded stars, the combined optical depth due to
the birth cloud (BC) and the inter-stellar medium (ISM) can be
several times larger than due to the ISM alone
\citep[e.g.][]{Charlot2000}. While most studies suggest that birth
clouds are transient features with lifetimes of a few tens of Myrs
\citep[e.g.][]{Blitz1980,Hartmann2001}, a recent work
\citep{Scoville2004} suggests that the low efficiency of star
formation in these clouds implies significantly longer lifetimes
($\sim 100$ Myrs). Recently formed stars may still migrate early
from such long-lived BCs, since the BCs feel hydro-dynamic forces
while the stars do not. However, the low gas density in early-type
galaxies imply lower hydro-dynamical forces, which could mean that
migration timescales are longer (Nick Scoville, private
communication).

In this study we apply a differential extinction scheme, in which
the extinction applied to flux from embedded stars is a factor
$\mu$ greater than the extinction felt by stars which are not
embedded - calculated from the mean H column density (see H2003,
Section 6.2). The key ingredients in this scheme are the factor
$\mu$ and the age limit $\tau_{max}$ of stars which are considered
to be embedded. Following the studies mentioned above, we
experiment with various values of $\tau_{max}$ in the range 10-100
Myrs and a range of values of $\mu$ between 3 and 10. We note that
\citet{Charlot2000} suggest $<\mu> \sim 3$, albeit with a large
scatter at lower and higher values - however this is mainly for
starbursting galaxies which have significantly different star
formation histories than the early-type galaxies in our sample
\citep[see also][]{Kong2004}.

A second issue regards the modelling of star formation itself.
Star formation is modelled as a continuous process proportional to
the mass of cold gas in the system - as long as there is gas to
fuel star formation, there are always some newborn i.e. zero-age
stars predicted in the model at \emph{any} epoch. However, the
question is whether, when galaxies are observed, \emph{true
zero-age stars} or indeed very young stars are actually seen. In
reality, star formation is unlikely to be an idealised continuous
process but probably occurs in random bursts of varying strengths
and durations. Galaxies are therefore observed sometime \emph{in
between} two starbursts i.e. there is a small \emph{time delay}
between a galaxy's most recent starburst and the point at which it
is observed. For example, NGC 205, an `actively' star-forming
dwarf elliptical
\citep[e.g.][]{Burstein1988,Wilcots1990,Dorman1995,Bertola1995} in
which the two most recent starbursts occurred $\sim$ 20 Myrs and
$\sim$ 500 Myrs ago \citep{Wilcots1990}, is noticeably bluer in
UV-optical colours than other elliptical galaxies
\citep{Burstein1988}. This suggests that it may not be correct to
model \emph{all} early-type galaxies with very young stars ($<20$
Myrs old).

If we assume that starbursts in real galaxies are randomly, i.e.
Poisson distributed, in time, then the distribution of time delays
between starbursts is exponential\footnote{This is a general
property of Poisson distributions}. The shape of the exponential
is determined by the number of starbursts per Gyr ($\kappa$). In
addition, the time lag distribution must be truncated at some
value $l_{max}$, since the time lag should generally be small and
could have a minimum value $l_{min}$ (e.g. $l_{min}=20$ Myrs and
$l_{max}=500$ Myrs, from the NGC 205 arguments above). In this
study we explore the ($\kappa$,$l_{min}$,$l_{max}$) parameter
space for values of $l_{min}$ in the range 0-0.05 Gyrs and
$l_{max}$ in the range 0.2-0.5 Gyrs. We assume that $\kappa = 10$
i.e. that a starburst has, on average, a timescale of $\sim$ 0.1
Gyrs, so that galaxies tend to have $\sim$ 10 bursts per Gyr.


\subsection{Reproducing the NUV CMR in the merger paradigm}
A one-to-one comparison between the model and observed photometry
requires that the two samples are consistent in terms of the
magnitude and redshift ranges covered by each. We therefore take
the following points into account while performing our
comparisons:

\begin{itemize}

    \item From Figure \ref{fig:blue_fraction} we see that, due to the magnitude limit of our
observed sample, luminosity function coverage is complete only for
galaxies brighter than $M(r)=-21.3$. We therefore restrict our
comparison \emph{only} to model galaxies with $M(r) < -21.3$.\\

    \item As mentioned before, the SDSS spectroscopic sample has a \emph{bright limit}
- bright galaxies which are very close do not have spectroscopy
and therefore do not appear in our observed sample. A large sample
of galaxies drawn from the SDSS DR3 indicates that this bright
limit lies around $M(r) = -22.7$ for $z < 0.05$. We therefore
remove such galaxies from the model sample.\\

    \item The GALEX detection limit ($m(NUV)<23$) implies that some
galaxies predicted in the model will not be `GALEX-detectable'.
The comparison must therefore include only model galaxies which
are GALEX-detectable.\\

    \item A fourth point which does not
particularly affect our comparison but is worth noting is that the
model itself has a completeness limit (H2003, Section 8.2 and
Table 3), because small haloes are not fully resolved in the
simulation due to the mass resolution of GALICS\footnote{In GALICS
the merging history is driven by the N-body \emph{backbone} on
which galaxies are `painted'. The mass resolution is therefore
determined by the mass of the dark matter particles ($\sim 10^{10}
M_{\odot}$). Other models,
\citep[e.g.][]{Benson2000,Khochfar2003}, use the Press-Schechter
formalism \citep{Press1974}, with theoretically infinite
resolution, so that the merging history of even the smallest
objects is fully resolved}. The completeness limit for GALICS in
\emph{r}-band is around -20.4. While the effects of such a lack of
resolution may propagate over the formal completeness limit of
-20.4, this does not affect our comparison if we restrict
ourselves to $M(r) < -21.3$, since this is well within the
completeness threshold.\\

    \item Finally, all model SEDs are redshifted to present day before
the comparison, to avoid uncertainties due to K-corrections. This
is important, because without robust low redshift templates for
the $UV$ spectra of early-type galaxies, we cannot reliably
estimate $UV$ K-corrections - we have already established that a
purely passively evolving template does not match at least 30
percent of the early-type galaxies at low redshifts. Later in this
section we present UV K-corrections based on the photometry of our
best-fit models.\\

\end{itemize}

\begin{figure}
\begin{center}
\includegraphics[width=3.5in]{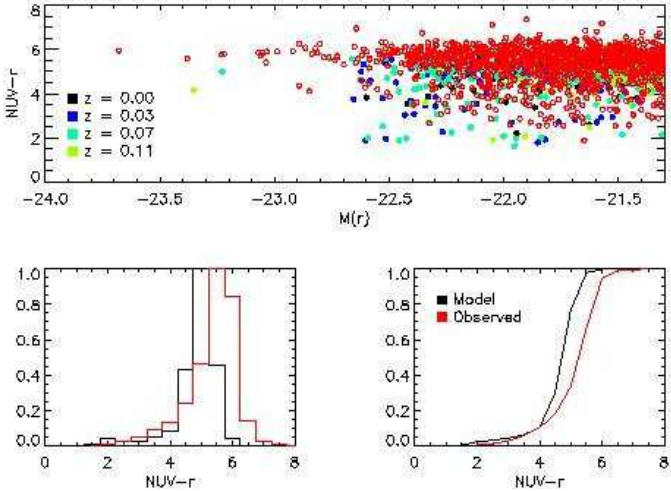}
\caption{Comparison of observed and model photometry in the case
where $\mu = 10$ and $\tau_{max} = 50$ Myrs, i.e. the BC + ISM
extinction is ten times that of ISM alone and BCs survive for 50
Myrs. The open red circles represent the observed photometry. All
other colours represent model galaxies at various redshifts (see
legend in plot).} \label{fig:mu10_tmax0.05}
\end{center}
\end{figure}

We generally find that reproducing the observed photometry without
invoking timelags is difficult. The assumption of moderate BC
survival spans ($\sim$10-50 Myrs) leaves the model photometry with
roughly correct scatter but with a zero-point which is
consistently too blue, regardless of how heavily attenuated the
flux from young stars is. To illustrate this point we show, in
Figure \ref{fig:mu10_tmax0.05}, the case where $\mu = 10$ and
$\tau_{max} = 50$ Myrs, i.e. the BC + ISM extinction is ten times
that of the ISM alone and BCs survive for 50 Myrs. Note that, for
clarity, in this and other similar figures that follow, we draw a
random sample of model galaxies which has the same size as the
observed sample. While the top panel of Figure
\ref{fig:mu10_tmax0.05} indicates that the scatter is reasonably
reproduced the colour distributions in the bottom panels clearly
show that the zero-point of model photometry is clearly too blue
by $\sim$ 0.5 mag in such a scenario.

\begin{figure}
\begin{center}
\includegraphics[width=3.5in]{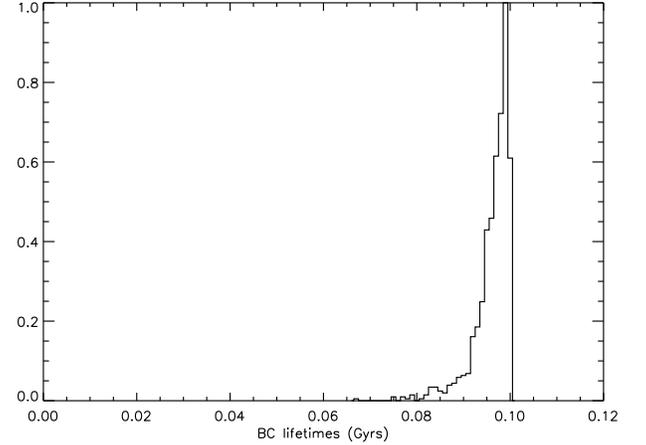}
\caption{Exponential distribution of BC lifetimes employed to
produce the comparison shown in Figure
\ref{fig:bcexpdist_250_10}.}
\label{fig:bc_lifetimes}
\end{center}
\end{figure}

\begin{figure}
\begin{center}
\includegraphics[width=3.5in]{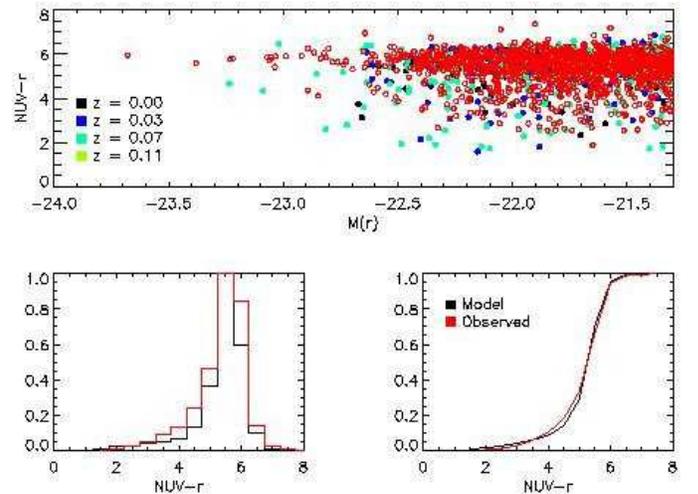}
\caption{Comparison of observed and model photometry in the case
where an exponential distribution of BC lifetimes is employed (see
Figure \ref{fig:bc_lifetimes}). The extinction assumed due to the
BCs is 5 times that due to the ISM alone i.e. $\mu = 5$. The open
red circles represent the observed photometry. All other colours
represent model galaxies at various redshifts (see legend in
plot).} \label{fig:bcexpdist_250_10}
\end{center}
\end{figure}

The situation can be significantly improved by invoking longer BC
lifetimes close to 100 Myrs - Figure \ref{fig:bcexpdist_250_10}
shows the comparison between the model and observed results when
an exponential BC lifetime distribution, highly skewed towards an
upper limit of 100 Myrs is used, along with a BC extinction of 5
times that due to the ISM alone. We find that this comparison
gives a reasonable fit to the photometry, although as mentioned
before, it is unclear whether recently formed stars within BCs
linger within them for the lifetime of the cloud itself.

\begin{figure}
\begin{center}
\includegraphics[width=3.5in]{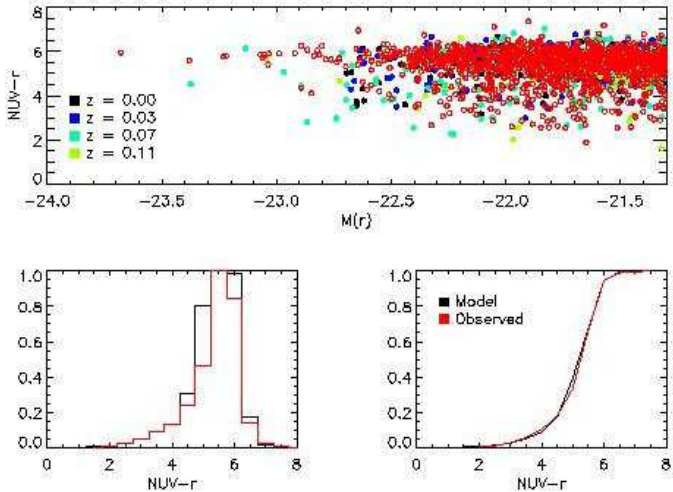}
\caption{Comparison of observed and model photometry in the
best-fit model where $\mu = 3$, $\tau_{max} = 30$ Myrs (c.f.
\citet{Charlot2000}, $l_{min}=20$ Myrs and $l_{max}=200$ Myrs. The
open red circles represent the observed photometry. All other
colours represent model galaxies at various redshifts (see legend
in plot).} \label{fig:best_fit_model}
\end{center}
\end{figure}

Invoking small timelags into the model photometry, \emph{while
keeping BC lifetimes short}, produces similar results. The
observed photometry can be reproduced best if we assume a minimum
timelag of 20 Myrs (c.f. NGC 205 arguments given above), a maximum
timelag of $\sim$ 200-300 Myrs and an exponential timelag
distribution constructed on the assumption that galaxies have, on
average, 10 starbursts per Gyr. We also assume that stars less
than 30 Myrs old reside in BCs and experience three times the
extinction due to the ISM alone (c.f. \citet{Charlot2000}. Using
the terminology of the previous section this translates to: $\mu =
3$, $\tau_{max} = 30$ Myrs, $\kappa = 10$, $l_{min} = 20$ Myrs and
$l_{max} = 200$ Myrs. Note that the BC + ISM extinction only
affects stars which are between 20 and 30 Myrs old since, by
construction, there are no stars younger than 20 Myrs in this
model (since $l_{min} = 20$ Myrs). This comparison is shown in
Figure \ref{fig:best_fit_model}. Henceforth, we refer to this
scenario as the `best-fit model'. As might be expected, the
goodness of fit is more sensitive to the minimum timelag $l_{min}$
- scenarios where $l_{min}< 20$ Myrs make both the zero-point too
blue and the scatter too large to fit the observations. The fit is
less sensitive to the maximum timelag but we use $l_{max}=200$
Myrs, the lowest value of $l_{max}$ which gives a good fit.

\begin{figure}
\begin{center}
\includegraphics[width=3.5in]{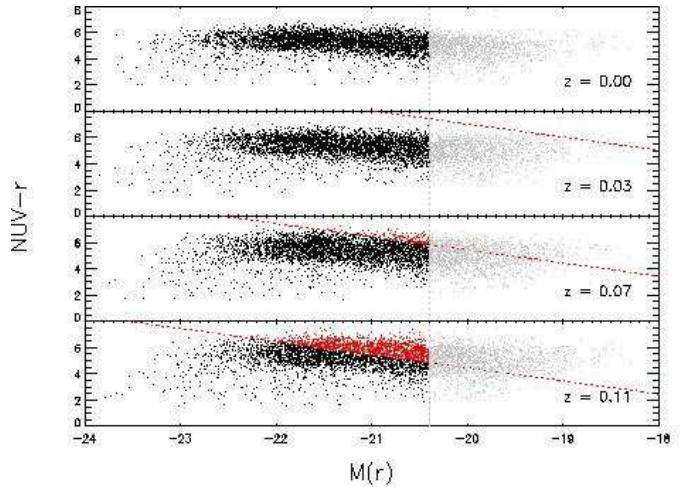}
\caption{Evolution of the $NUV$ CMR in the best-fit model. The red
dots indicate galaxies which are below the detection threshold of
GALEX. The grey dots lie in the region below the completeness
limit of the GALICS model. The detection limit encroaches severely
on the model sample at the upper end of our redshift range ($z =
0.11$).} \label{fig:bestfit_nuvcmr}
\end{center}
\end{figure}

As the bottom panels of Figure \ref{fig:best_fit_model} indicate,
the best-fit scenario reproduces the colour distribution of the
observed sample well. The evolution of the (complete) $NUV$ CMR in
this scenario is shown in Figure \ref{fig:bestfit_nuvcmr}. Note
that the number of galaxies in each redshift bin is around $\sim$
6000. The red dots indicate galaxies which are below the detection
limit of GALEX i.e. they are not GALEX-detectable. The grey dots
lie in the region below the completeness limit of the GALICS
model. The detection limit encroaches severely on the model sample
at the upper end of our redshift range ($z = 0.11$).

\begin{figure}
\begin{center}
\includegraphics[width=3.5in]{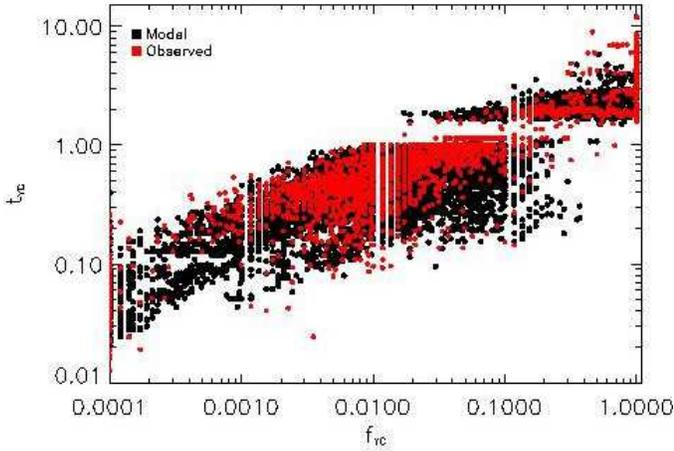}
\caption{The two-component method applied to model early-types in
the GALICS hierarchical merger model which is used in this study.
Black points represent model galaxies and the red points represent
the observations. Note that this comparison is not restricted to
galaxies with $M(r)<-21.3$. All observed galaxies and all model
galaxies above the completeness limit of $M(r) = -20.4$ are
plotted. There are no significant systematic differences between
the positions of model and observed galaxies in this parameter
space, implying that the observed photometry across the UV and
optical spectral ranges is reproduced reasonably well by the
models.} \label{fig:ts_compare}
\end{center}
\end{figure}

While it is apparent that the best-fit model reproduces the $UV$
photometry of the observed sample, we now apply the two-component
analysis used in Section 3 to the predicted photometry from the
best-fit model. The position of an object in $(f_{YC},t_{YC})$
space is a direct function of the \emph{shape} of its SED, which
in turn drives its photometric properties both in the $UV$ and the
optical spectral ranges. A comparison, in this parameter space,
between the positions of the model galaxies and the observed
population indicates the overall robustness of the fit, not only
in terms of the $NUV$ colour but in terms of the shape of the
complete SED. The derivation of $(f_{YC},t_{YC})$ for model
galaxies uses the same process as that used for the observed
sample. We show this comparison in Figure \ref{fig:ts_compare}.
Note that we show all available galaxies in the observed sample
and similarly all model galaxies above the GALICS completeness
limit - we do not restrict the analysis to galaxies with
$M(r)<-21.3$. We find that there are no systematic differences
between the positions of model and observed galaxies in this
parameter space, which implies that the observed photometry across
the $UV$ and optical spectral ranges is reproduced reasonably well
by the best-fit model. Table \ref{tab:SAM_rsf_properties}
summarises the properties of the recent star formation in the
galaxies in the best-fit model. On average, low-redshift
early-type galaxies are predicted to have $\sim$ 1.5-2.5 percent
of their stellar mass formed within the last Gyr. The
mass-weighted age of this star formation is between 0.4 and 0.6
Gyrs and its luminosity-weighted average age is around 0.5 Gyrs.

\begin{deluxetable}{ccccc}

\tablecaption{The average properties of recent star formation in
model early-types.\label{tab:SAM_rsf_properties}}

\tablehead{\colhead{z} & \colhead{$<f>$} & \colhead{$<\tau_{m}>$}
& \colhead{$<\tau_{v}>$} & \colhead{$<\tau_{r}>$}} \startdata

   0.03  &  0.014  &  0.61  &  0.55  &  0.55\\
   0.07  &  0.021  &  0.62  &  0.55  &  0.57\\
   0.11  &  0.025  &  0.50  &  0.37  &  0.38\\

\tablecomments{$<f>$ is the average `RSF' mass fraction i.e. the
stellar mass fraction within the last Gyr. $<\tau_{x}>$ is the
average age of this RSF fraction, in Gyrs, weighted by the
quantity indicated in the subscript $x$. Subscript $m$ indicates
mass, $v$ indicates $V$-band luminosity and $r$ indicates r-band
luminosity.}
\enddata
\end{deluxetable}

\begin{figure}
\begin{center}
\includegraphics[width=3.5in]{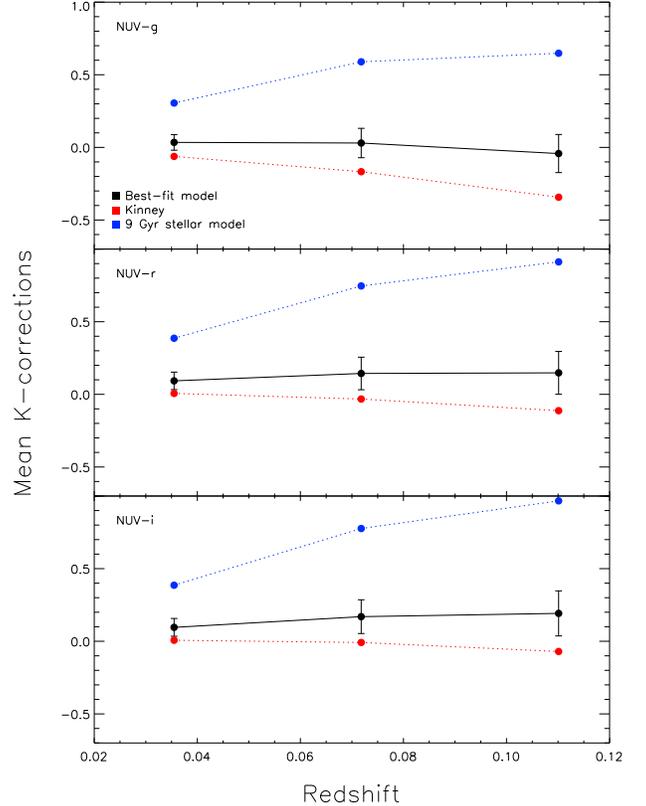}
\caption{K-corrections for $UV$ colours from our best-fit model in
the redshift range spanned by our observed sample (black). Also
shown are K-corrections obtained from a passively evolving 9 Gyrs
old stellar model (blue) and the elliptical template from the
Kinney spectral atlas of galaxies (red).} \label{fig:kcorr_galics}
\end{center}
\end{figure}

We end this section by computing K-corrections \citep{Hogg2002}
from our best-fit models. As suggested by our previous analysis,
purely old, passively evolving populations are clearly \emph{not}
adequately good templates for the $UV$ spectra of early-type
galaxies, even though the $UV$ emission in these galaxies is weak.
In Figure \ref{fig:kcorr_galics}, we show K-corrections for $UV$
colours from our best-fit model in the redshift range spanned by
our observed sample (black dots with error bars). We also compare
these K-corrections to those obtained from a passively evolving 9
Gyrs old SSP (blue) and the elliptical template from the Kinney
spectral atlas of galaxies \citep[red,][]{Kinney1996}. The use of
a 9 Gyr SSP is motivated by the fact that Bernardi et al. (2003d)
find that their early-type sample can be best-fit by a 9 Gyr
passively evolving SSP, formed at high redshift.

\subsection{The role of mergers and progenitor bias}
\begin{figure}
$\begin{array}{c}
\includegraphics[width=3.5in]{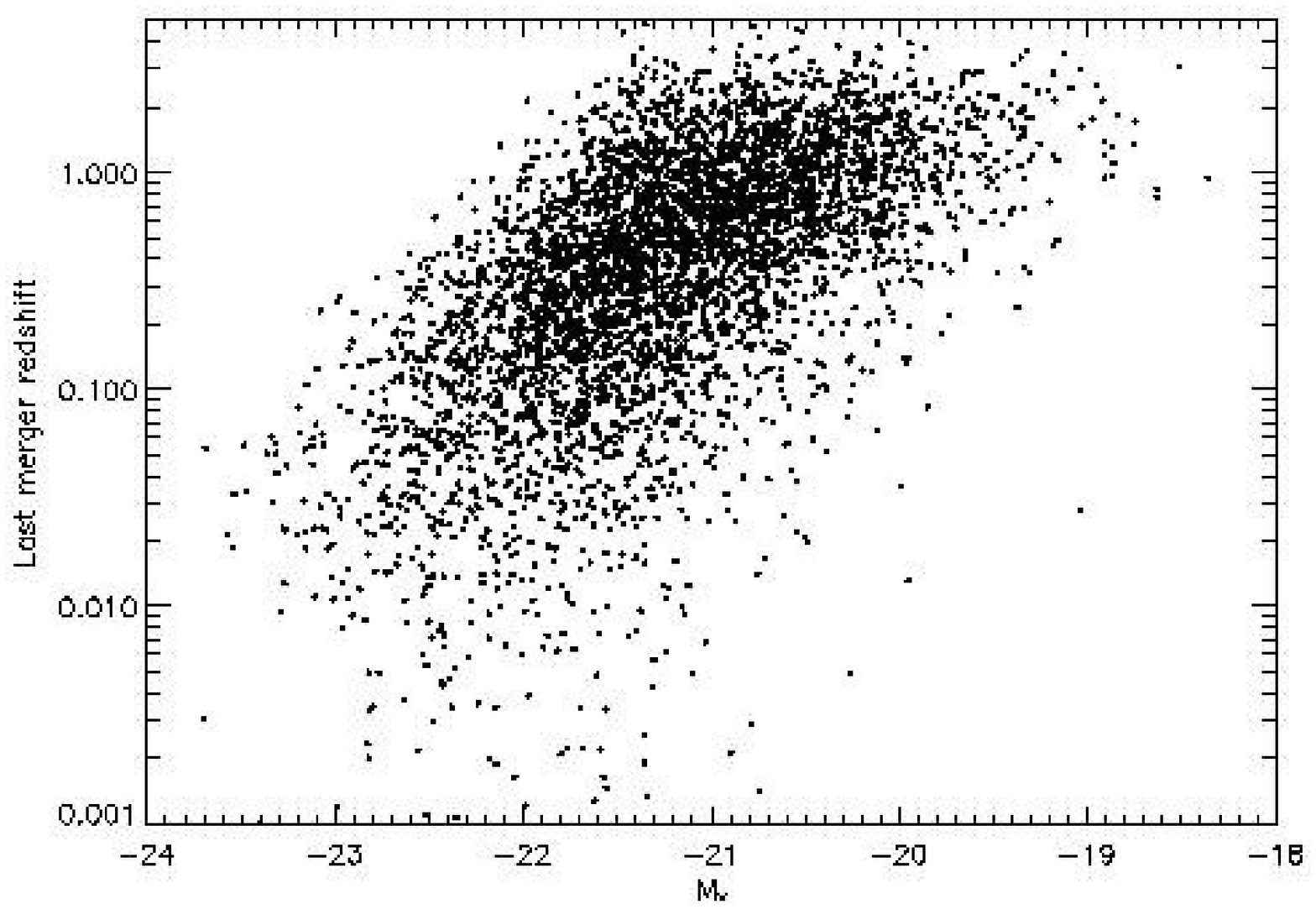}\\
\includegraphics[width=3.5in]{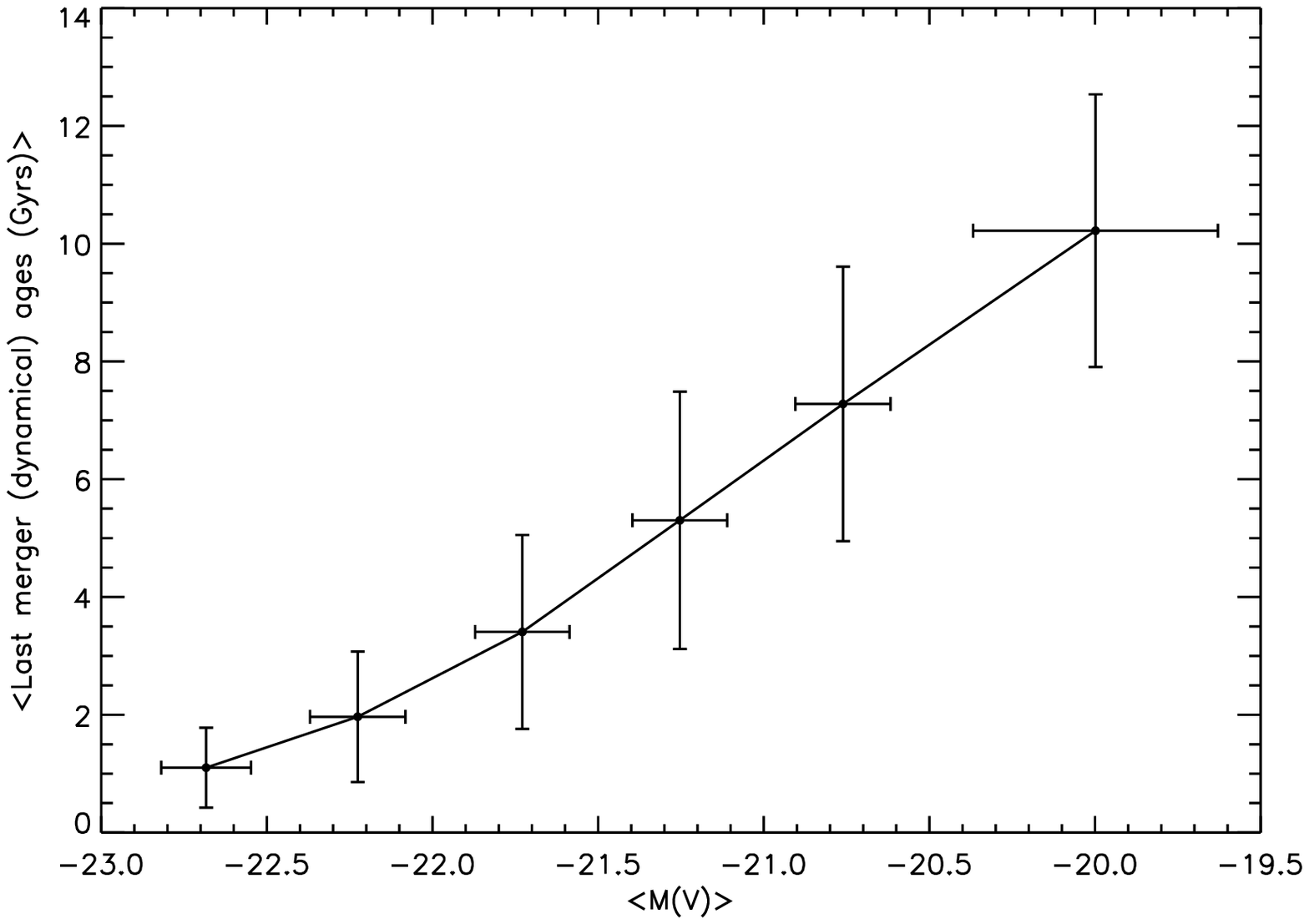}
\end{array}$
\caption{TOP: Redshifts of most recent mergers experienced by
early-type galaxies at $z=0$. These are essentially the
\emph{dynamical ages} of these galaxies i.e. the epoch at which
the galaxy in its present (morphological) form was created.
BOTTOM: Average dynamical ages of galaxies as a function of their
luminosity (size). As expected, in a hierarchical picture of
galaxy formation, larger galaxies are assembled later i.e. have
smaller dynamical ages, although the ages of their stellar
populations are, on average, older \citep{Kaviraj2005a}.}
\label{fig:last_mergers}
\end{figure}

Star formation in the merger paradigm proceeds through one of two
channels - \emph{quiescent} star formation governed by the mass of
available cold gas and the dynamical timescale of the galaxy
(H2003, Section 4.1) or \emph{merger driven} star formation as a
result of dynamical interactions between galaxies (H2003, Section
5). As Figure \ref{fig:last_mergers} indicates, it is possible for
galaxies to have experienced very recent mergers, within the last
Gyr of look-back time. The top panel in Figure
\ref{fig:last_mergers} shows the redshifts of the \emph{most
recent} mergers experienced by early-type galaxies at $z=0$. These
are essentially the \emph{dynamical ages} of these galaxies i.e.
the epoch at which the galaxy in its present (morphological) form
was created. The bottom panel shows the average trend in dynamical
ages as a function of luminosity (size). In a hierarchical picture
of galaxy formation larger early-types are assembled later, i.e.
have smaller dynamical ages, although the ages of their stellar
populations are, on average, older \citep{Kaviraj2005a}.

\begin{figure}
\includegraphics[width=3.5in]{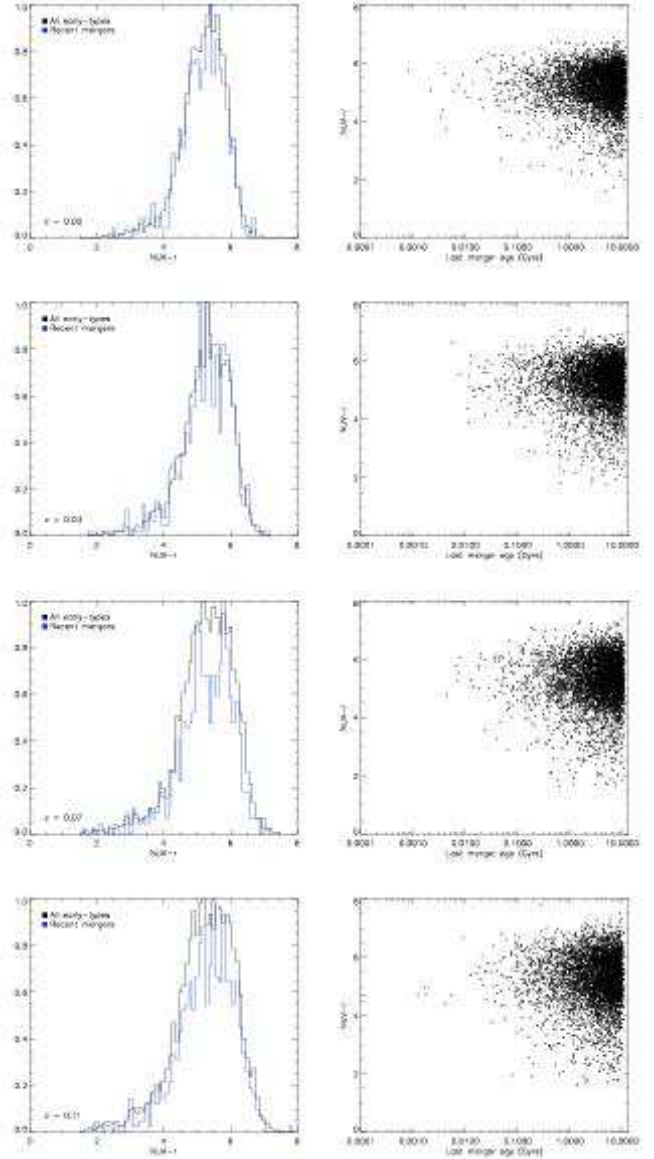}
\caption{LEFT: Comparison between the $UV$ colours of early-type
galaxies which have had recent mergers (blue) within the last Gyr
of look-back time with the colour distribution of all galaxies.
RIGHT: The $UV$ colours of early-type galaxies as a function of
its dynamical (last-merger) age. The dynamical age is explained in
the text and in the caption to Figure \ref{fig:last_mergers}. We
find that recent mergers do not have a preponderance of blue
galaxies, suggesting that mergers at low redshift are
predominantly \emph{dry} and not the cause for the blue scatter in
the UV colours of early-type galaxies.}
\label{fig:merger_colours}
\end{figure}

We investigate the impact such recent mergers may have on the
colour of the early-type remnants, and the extent to which they
may affect the blue colours exhibited by some early-type galaxies.
In Figure \ref{fig:merger_colours}, we compare the $UV$ colour
distribution of galaxies which have experienced recent mergers
(within the last Gyr), with the global colour distribution of all
early-type galaxies. We find that recent mergers do not
have a preponderance of blue galaxies, suggesting that {\em mergers at
low redshift are predominantly dry and not the cause for
the blue scatter} in the $UV$ colours of early-type galaxies - star
formation from the quiescent mode is at least as important as that
due to mergers.


We finish this section by briefly mentioning the issue of
\emph{progenitor bias}, which affects all studies of early-type
galaxies at $z>0$, in the framework of the merger paradigm.
Fundamental to merger-driven morphological transformations is the
implication that, at higher redshifts, a progressively larger
fraction of mass that eventually resides in present-day
early-types is potentially locked up in late-type systems
\citep{VD2001}. Like all other early-type studies, this
investigation has attempted to trace the evolution of the
\emph{local} early-type population by looking \emph{only at their
early-type progenitors in the redshift range $0<z<0.11$} - any non
early-type progenitors have been excluded, for the simple reason
that it is not possible to easily identify them! Although we
expect progenitor bias to be smallest at low redshifts (indeed
this is the big advantage of studying \emph{nearby} early-type
galaxies!), it is instructive to quantify this effect and gauge
its possible impact on the blue fraction derived in Section 4.

\begin{figure}
$\begin{array}{c}
\includegraphics[width=3.5in]{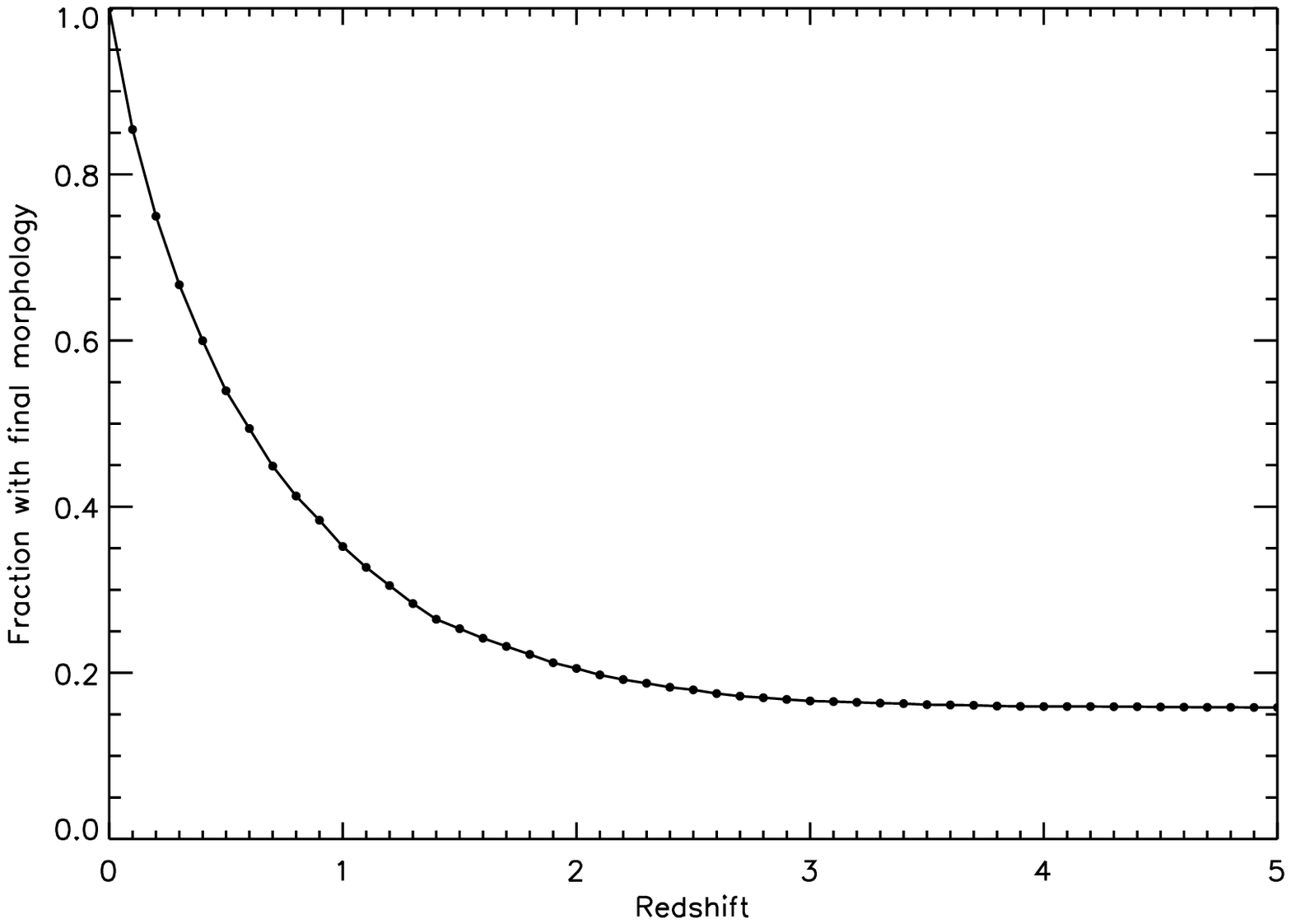}\\
\includegraphics[width=3.5in]{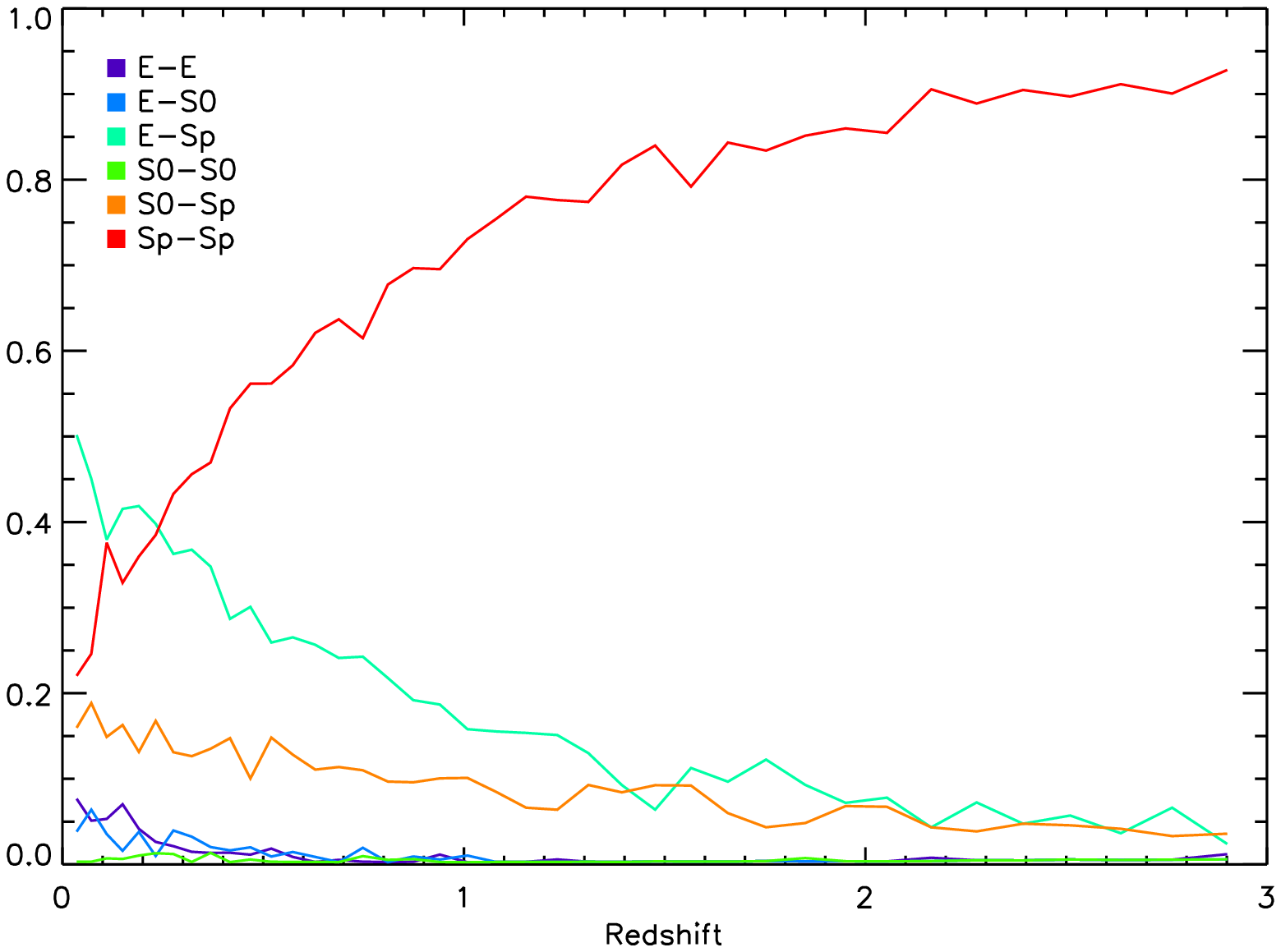}\\
\end{array}$
\caption{TOP: The fraction of local ($z=0$) early-type galaxies
which have already attained their final (early-type) morphology,
i.e. had their last merger, as a function of redshift. BOTTOM:
Progenitor types in binary mergers. At low redshift mergers have a
higher probability of having at least one early-type progenitor.
See \citet{Khochfar2003} for a detailed analysis of the progenitor
morphologies of early-type galaxies.} \label{fig:progenitor_bias}
\end{figure}

The top panel in Figure \ref{fig:progenitor_bias} indicates the
extent of progenitor bias by plotting the fraction of the $z=0$
early-type galaxies which have already undergone their last merger
and assumed their final early-type morphology as a function of
redshift. At $z \sim 0.1$, roughly 10 percent of present-day
early-type galaxies are still `in pieces'. If, for example, these
`pieces' were blue late type systems, then we could have
underestimated the blue fraction in Section 4. However, we have
already shown (Figure \ref{fig:merger_colours}) that recent
mergers in this redshift range have a similar UV colour
distribution to the rest of the population i.e. that recent
mergers are not preferentially bluer. Indeed, mergers at low
redshift have a high probability of involving at least one
early-type progenitor (bottom panel of Figure
\ref{fig:progenitor_bias}). Note that our results are consistent
with \citet{Khochfar2003} who were the first to present a detailed
analysis of the progenitor morphologies of early-type galaxies.
Thus, since progenitor bias is small within our redshift range and
mergers tend to be dry, neglecting non-early-type progenitors will
not affect the blue fraction derived in this study.


\section{Revisiting the monolithic hypothesis}
The monolithic hypothesis remains the simplest and perhaps most
elegant explanation for a remarkably wide range of properties of
large early-type galaxies \citep[e.g.][]{Peebles2002}. Indeed, a
careful treatment using N-body/SPH simulations
\citep{Chiosi2002,Tantalo2002} suggests that monolithic formation
can reproduce a wide spectrum of chemo-photometric properties of
large early-types. Until now these properties have been restricted
to the optical spectrum - the $UV$ photometry presented in this
paper provides, for the first time, a new constraint on the
formation of such large early-type systems. While it is apparent
that a merger-based treatment is able to reproduce the $UV$ +
optical photometry of local early-types fairly accurately, the RSF
mass fractions ($\sim$ 1 to 3 percent) involved are small. The
presence of young stars in these galaxies indicates that there is
fuel, i.e. cold gas, available for star formation at low redshift.
It is, therefore, natural to ask whether the required level of
fuel \emph{could} be supplied in a traditional monolithic scheme.

\begin{figure*}
$\begin{array}{cc}
\includegraphics[width=3.5in]{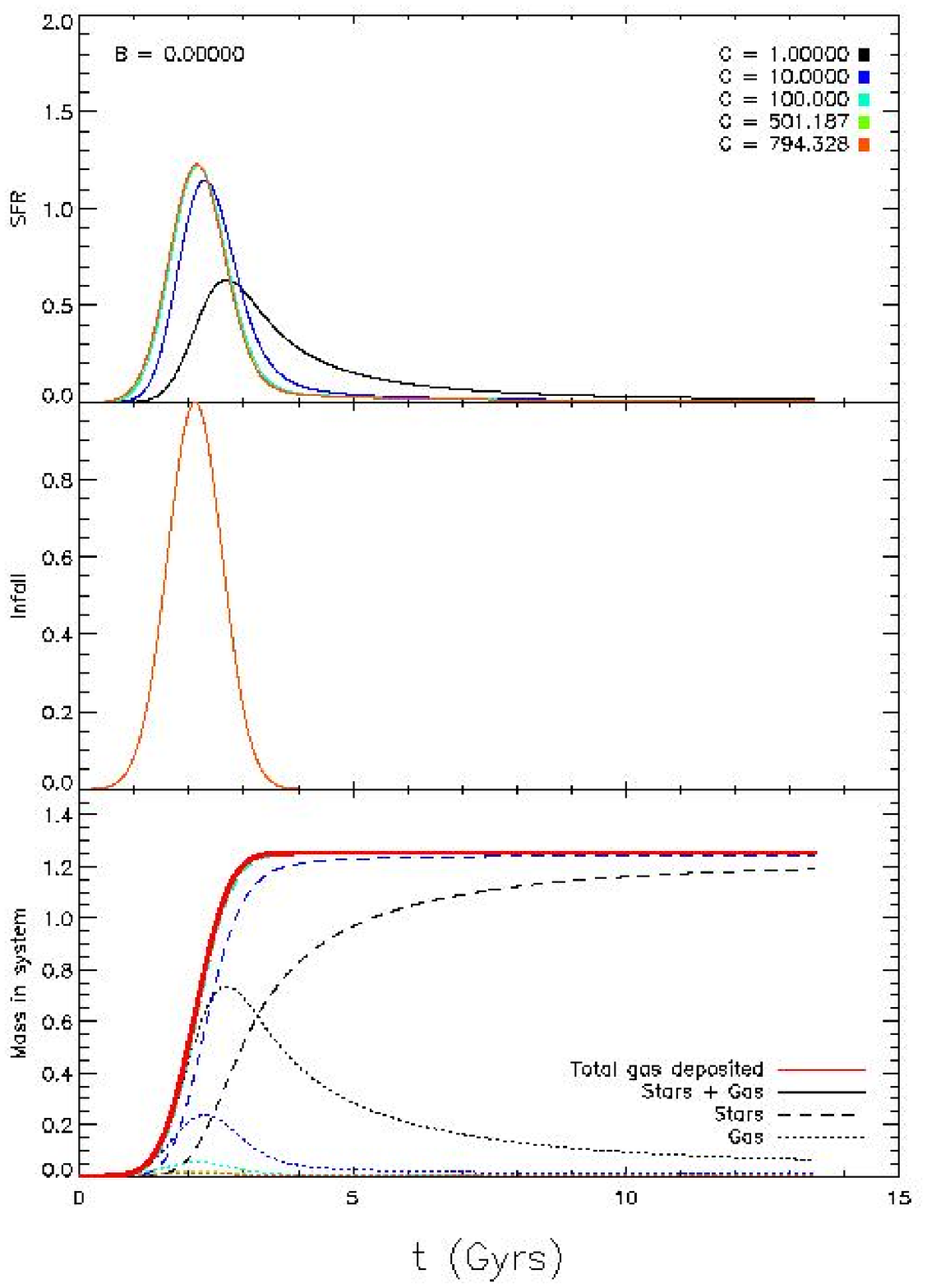} &
\includegraphics[width=3.5in]{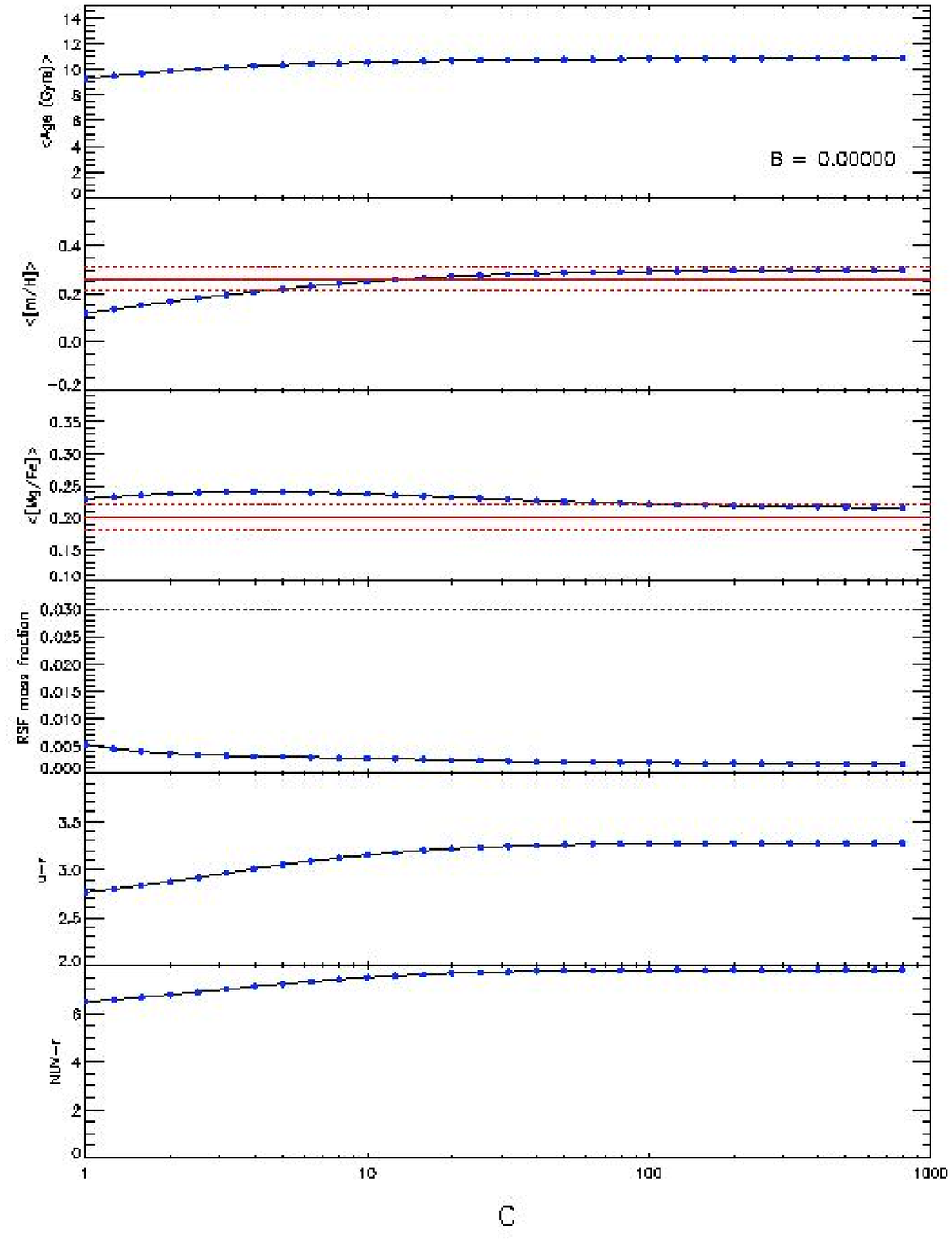}\\
\end{array}$
\caption{LEFT-HAND COLUMN: Star formation rate (top row), gas
infall (middle row) and the evolution of star and gas (bottom row)
for $B$ fixed at 0 and various values of $C$ (see legend in the
top row). RIGHT-HAND COLUMN: Average quantities - rows from top to
bottom show average age, metallicity, alpha-enhancement, fraction
of stars formed within the last Gyr (the `RSF' fraction), $u-r$
colour and $NUV-r$ colour respectively, as a function of $C$ (with
$B$ fixed at 0). Also indicated, in the relevant rows, are the
average values of [m/H] and [Mg/Fe] computed by
\citet{Trager2000a} for local early-type galaxies.}
\label{fig:beq0}
\end{figure*}

\begin{figure*}
$\begin{array}{cc}
\includegraphics[width=3.5in]{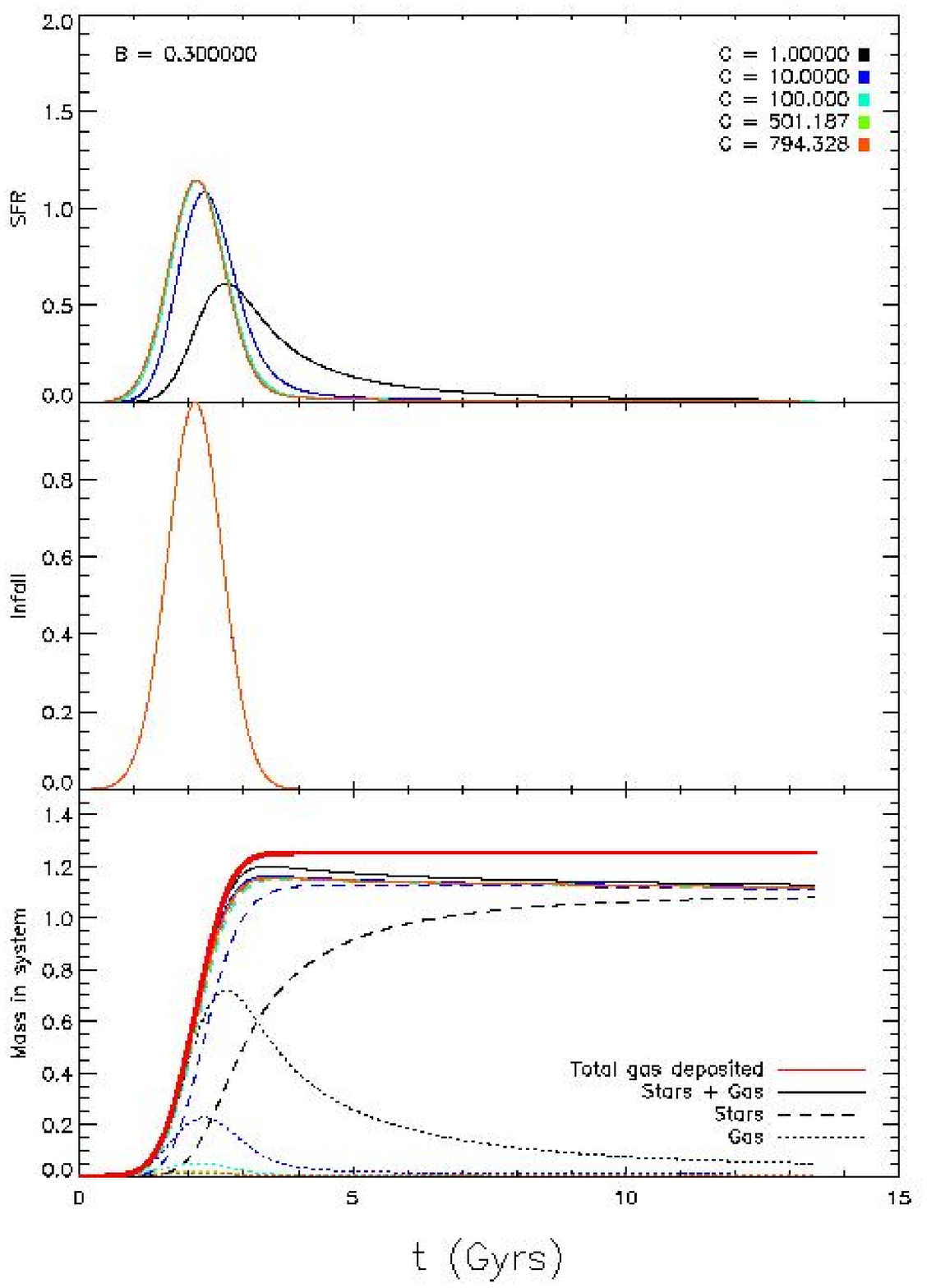} &
\includegraphics[width=3.5in]{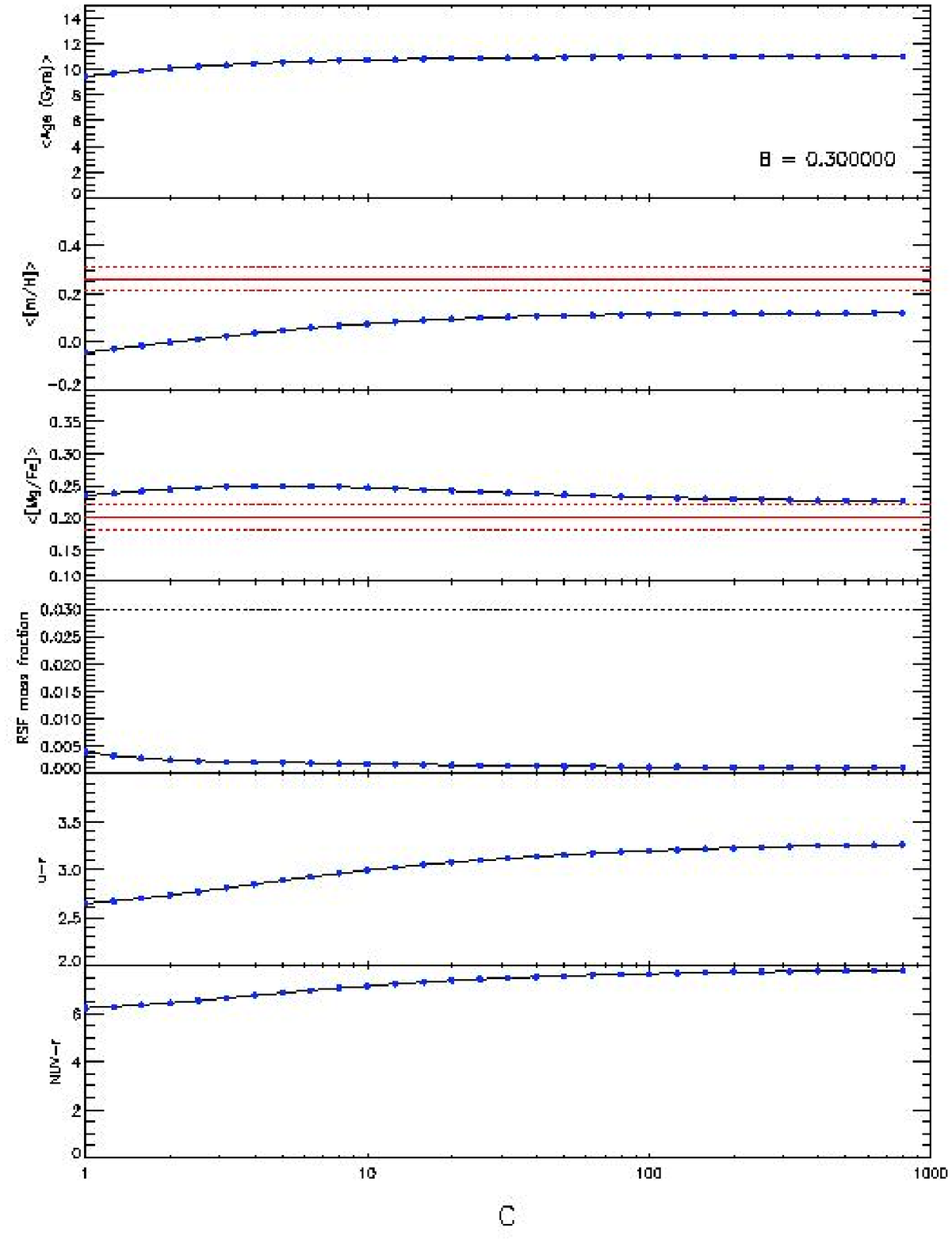}
\end{array}$
\caption{Same as Figure \ref{fig:beq0} except for $B=0.3$.
LEFT-HAND COLUMN: Star formation rate (top row), gas infall
(middle row) and the evolution of star and gas (bottom row) for
$B$ fixed at 0 and various values of $C$ (see legend in the top
row). RIGHT-HAND COLUMN: Average quantities - rows from top to
bottom show average age, metallicity, alpha-enhancement, fraction
of stars formed within the last Gyr (the `RSF' fraction), $u-r$
colour and $NUV-r$ colour respectively, as a function of $C$ (with
$B$ fixed at 0.3). Also indicated, in the relevant rows, are the
average values of [m/H] and [Mg/Fe] computed by
\citet{Trager2000a} for local early-type galaxies.}
\label{fig:beq0.3}
\end{figure*}

\begin{figure*}
$\begin{array}{cc}
\includegraphics[width=3.5in]{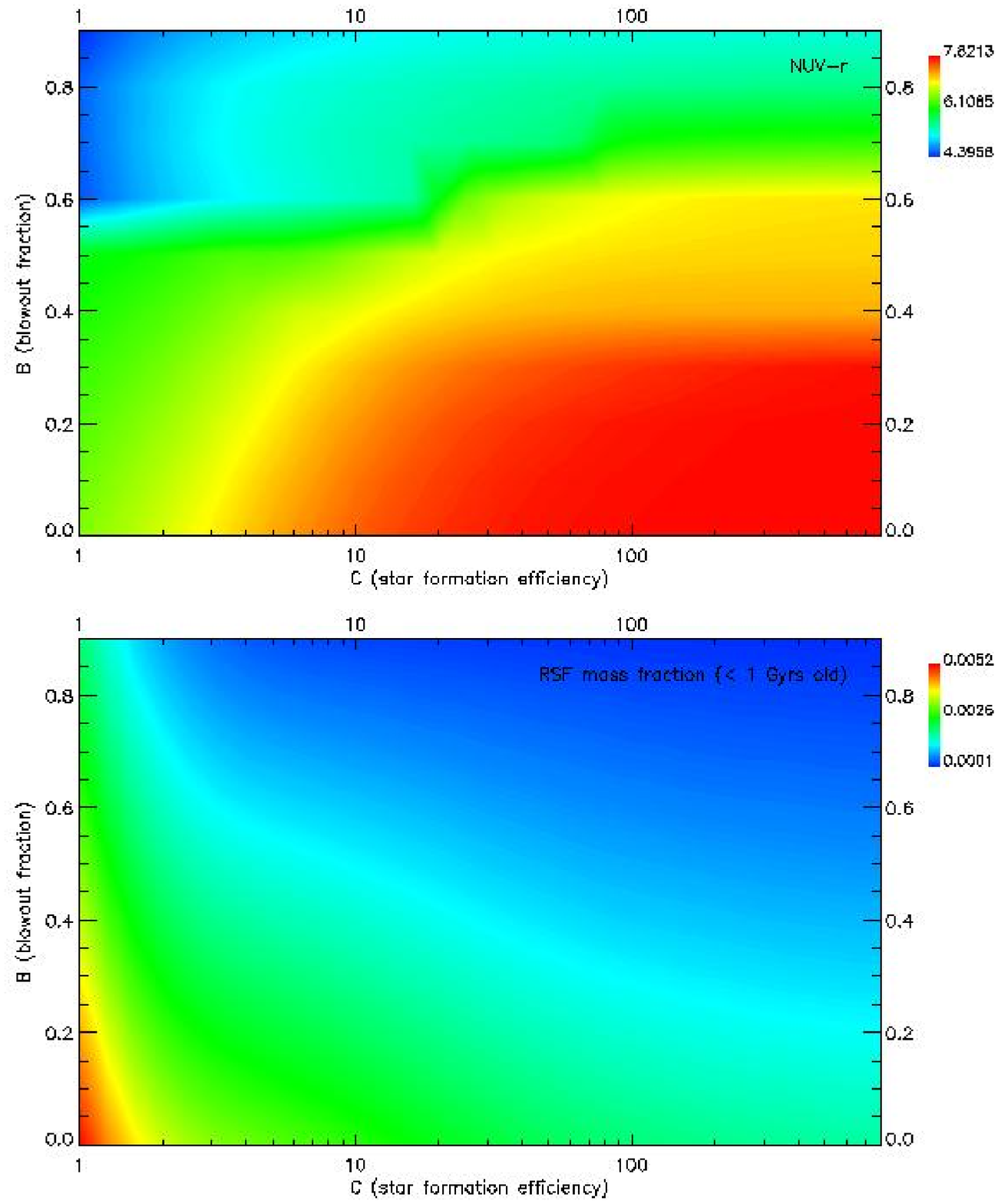} &
\includegraphics[width=3.5in]{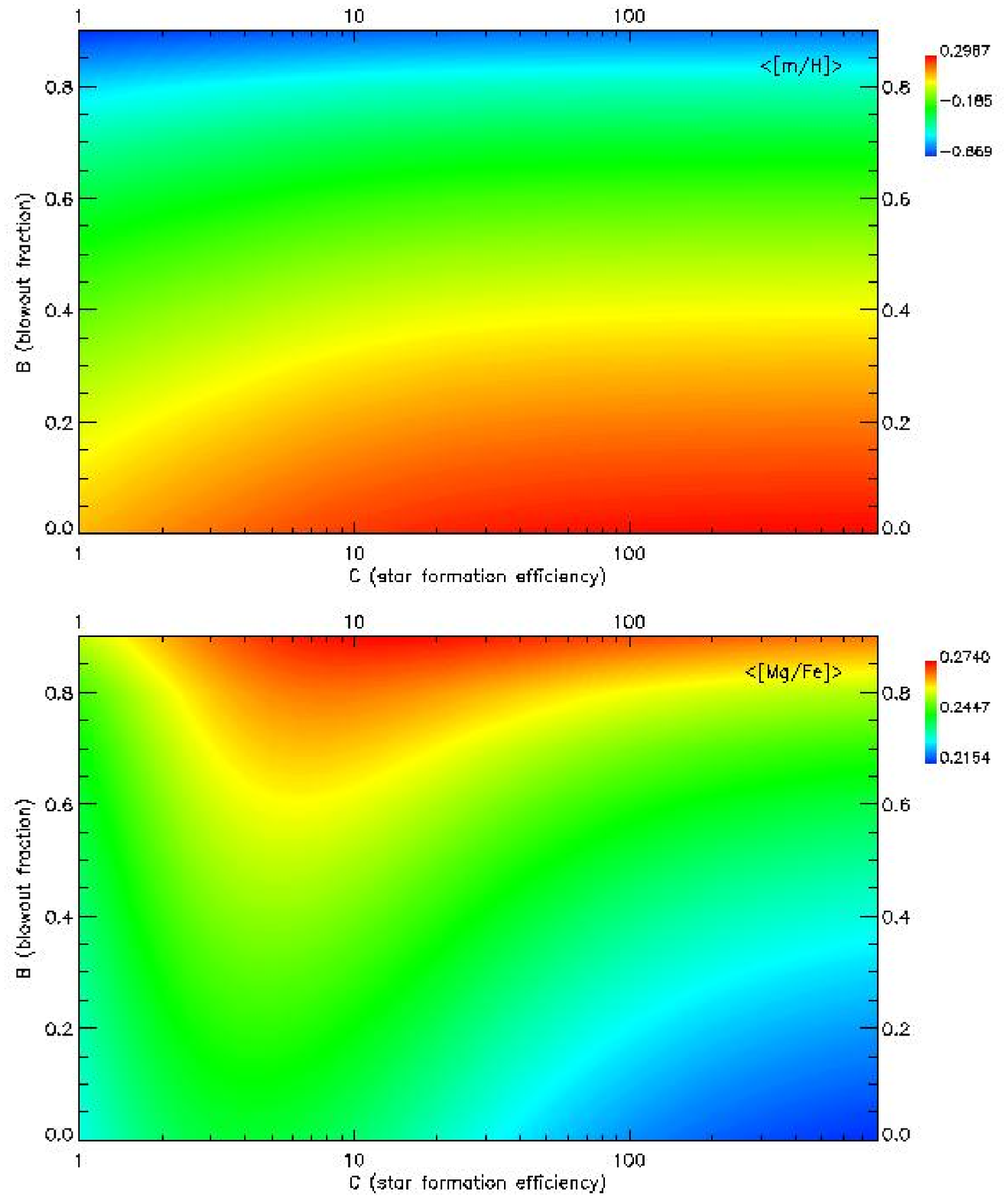}
\end{array}$
\caption{Summary of average quantities as a function of $B$ and
$C$: $NUV-r$ colour (top left), RSF fraction (bottom left), mean
metallicity (top right) and mean alpha-enhancement (bottom
right).} \label{fig:bc_summary}
\end{figure*}

Monolithic models propose a high star formation efficiency at high
redshift, which drives a violent episode of star formation
resulting in the bulk of the galaxy forming within a short
timescale at high redshift (e.g. $z<2$). However, stars created in
the primordial burst would \emph{recycle} a fraction of their mass
back into the ISM through stellar winds and supernova ejecta. A
fraction or all of this internally recycled gas \emph{could} fuel
further star formation, without invoking any external gas
accretion or interactions. The aim of this section is to
investigate whether, given reasonable assumptions for the
evolution and chemical enrichment of such a monolithic system, we
can reproduce the RSF and the observed $UV$ photometry of large
early-type systems \emph{that form part of the blue fraction}
\footnote{While we could simply explore the \emph{amount} of
recycled material expected in a typical monolithic simulation, it is
necessary to simulate the accompanying chemical enrichment because
the \emph{NUV-r} colour is a function of both the total amount and
the metallicity of the recycled gas}. The reason for restricting
ourselves to large blue galaxies is because, as mentioned before,
red galaxies \emph{can} be reconciled with a monolithic scenario,
especially if we assume that part of the scatter in the red
sequence is determined by varying levels of intrinsic dust in
these galaxies. This monolithic analysis is useful because, if low
level star formation seen in the observed blue population can be
produced simply by returned material from stars formed in the
primordial burst, then there would, in principle, be no need to
invoke a merger based scenario to produce early-type galaxies!

\subsection{Simulating `monolithic collapse'}
The model used here \citep[see][]{Ferreras2000b} follows the
standard chemical enrichment equations
\citep[e.g.][]{Tinsley1980}. Stellar yields are taken from
\citet{Thielemann1996} for stellar masses $M_{*}>10M_{\odot}$ and
from \citet{Vandenhoek1997} for lower mass stars. We use a
standard Salpeter IMF and note that assume
instantaneous recycling but \emph{not} instantaneous mass loss, so
that the lifetime of stars of various masses is taken into
account.

We simulate a scenario where (gaussian) gas infall, with a short
timescale ($\sim$ 1 Gyr), occurs at `high' redshift. Most studies
which support the monolithic hypothesis agree on a formation epoch
somewhere before $z=2$ - it is difficult to pinpoint it any
further using the available data. Our conclusions are not
sensitive to the exact redshift of formation so we use a fiducial
value of $z=3$. The `formation redshift' is taken as the epoch at
which the gas infall \emph{peaks}. The star formation rate is
governed by the generally accepted Schmidt law \citep{Schmidt59}:
\begin{equation}
\Psi = Cg^\alpha
\label{eq:sfr}
\end{equation}
where $\Psi$ is the star formation rate, $C$ is the star formation
efficiency, $g$ is the gas mass and $\alpha=1.5$.

We explore the behaviour of the system using two free parameters.
The first and most significant free parameter is the star
formation efficiency $C$ in Equation \ref{eq:sfr}. Low values of
$C$ result in more extended star formation, while high values of
$C$ result in faster consumption of gas and a shorter episode of
star formation - note that $C$ is not numerically restricted to a
value between 0 and 1. As expected, increasing $C$ to arbitrarily
high values does not result in the star formation increasing
without limit, simply because the system cannot consume gas faster
than it is being put in! We explore the behaviour of the system in
this `limiting' high $C$ regime, where star formation essentially
tracks the gas infall very closely i.e. as soon as gas becomes
available, it is almost instantly converted to stars. As a result,
the overwhelming bulk of the galaxy is already constructed by the
time gas infall ends at high redshift - mimicking the fundamental
feature of a `monolithic collapse'. However, star formation does
not stop when gas infall ceases. Gaseous ejecta returned from
dying stars is gradually released back into the ISM over a Hubble
time, providing potential fuel for further star formation. It is
the quantity of this recycled gas that we investigate here - in
particular, we explore whether there is \emph{enough} gas in this
recycled material to fuel late stage star formation and reproduce
the RSF mass fractions seen in large blue early-types in our
observed sample.

The second free parameter is a simple prescription for galactic
winds, which is parametrised by assuming that a certain fraction
($B$) of gaseous ejecta is permanently lost from the system. $B$
varies between 0 and 1. We are primarily interested in scenarios
where $B$ is small, since we are exploring large early-type
galaxies which should have potential wells deep enough to retain
most of the ejected gas. $B=0$ corresponds to essentially
closed-box evolution after the initial gas infall stage.

\subsection{RSF in `monolithic' early-type galaxies}
We begin by looking at the simplest scenario, where $B=0$, i.e.
all ejecta are retained by the system - this is also, of course,
the scenario which \emph{maximises} the late stage returned gas
fraction. The $B=0$ scenario is summarised in Figure
\ref{fig:beq0}. The left-hand column shows the star formation rate
(top row), gas infall (middle row) and the evolution of stars and
gas (bottom row) for various values of $C$ (see legend in the top
row), with $B$ fixed at 0. We note that the `limiting' high $C$
regime, which mimics monolithic collapse, is achieved for values
of $C$ above $\sim$ 100. Above this value, increasing $C$ does not
affect the SFR (top row), because the SFR is effectively limited
by the rate of gas infall, since stars cannot be produced faster
than gas is being deposited in the system. As mentioned before,
increasing $C$ allows the SFR to track the gas infall function
more closely, as can be seen in Figure \ref{fig:beq0} - the yellow
SFR curve, which represents a system with high $C$, peaks almost
coevally with the gas infall, whereas the black SFR curve, which
represents the lower $C$ regime, produces more extended star
formation and has its peak significantly displaced from the peak
of the gas infall. Our focus in this section will be exclusively
on the high $C$ regime ($C>100$), in which star formation after
the initial gas infall stage is fuelled \emph{only} by recycled
gas.

The right-hand column in Figure \ref{fig:beq0} shows average
quantities produced by such a scenario. The rows from top to
bottom show the average age, average metallicity, average
alpha-enhancement, fraction of stars formed within the last Gyr
(the `RSF' fraction), $u-r$ colour and $NUV-r$ colour
respectively, as a function of $C$ (with $B$ fixed at 0). We also
indicate, in the relevant rows, the average values (and their
formal errors) of [m/H] and [Mg/Fe] computed by
\citet{Trager2000a} for local early-type galaxies \footnote{Note
that \citet{Trager2000a} find milder values of
$\langle$[m/H]$\rangle$ and $\langle$[Mg/Fe]$\rangle$ than in
previous studies
\citep[e.g.][]{Weiss1995,Trager1997,Greggio1997}}. Figure
\ref{fig:beq0} indicates that, in the $B=0$ scenario, `monolithic'
values of $C$ ($C>100$) predict present-day RSF fractions of
$<0.005$. In comparison, we find that large ($>L_*)$, blue
($NUV-r<5.5$) galaxies in the merger models have present-day RSF
fractions of $\sim 0.03$. The monolithic RSF fraction, driven
purely by the recycled material from stars, is generally a factor
5-6 lower than what is required, in a merger framework, to
reproduce the colours of large, blue early-type galaxies in the
observed sample.

It is instructive to explore the full range of values for $B$,
although our exclusive interest in large galaxies implies that
values of $B$ should be low. As we show below, very high values of
$B$ result in excessive metal loss producing galaxies which are
too metal-poor to fit the observed metallicities of local
early-types. To illustrate the trend in $B$ we show, in Figure
\ref{fig:beq0.3}, the case where $B=0.3$ i.e. 30 percent of
stellar ejecta are permanently lost from the galaxy. We find, in
this case, that galaxies are generally more metal-poor than the
mean relations derived by \citet{Trager2000a} - increasing $B$
simply exacerbates this situation.

\begin{figure}
\includegraphics[width=3.5in]{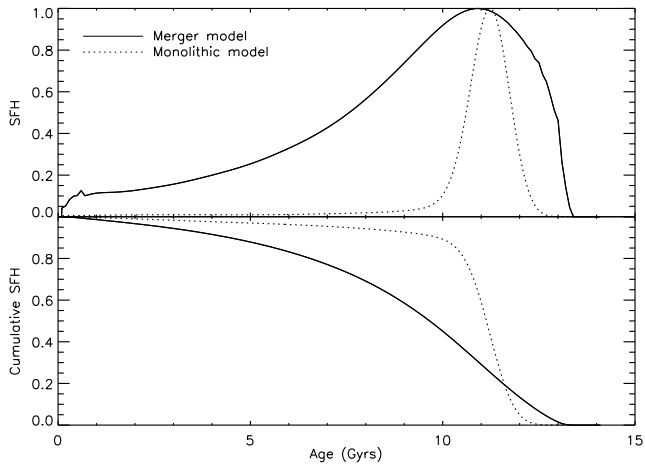}
\caption{Comparison between a `monolithic' SFH and the average SFH
of large ($>L_*)$, blue ($NUV-r<5.5$) early-type galaxies
predicted by the merger model (which fit the colours of
corresponding observed early-type galaxies). The top panel shows
the non-cumulative SFH while the bottom panels compares the
cumulative SFH.} \label{fig:compare_sam_mono}
\end{figure}

We summarise our investigation of monolithic models in Figure
\ref{fig:bc_summary}, which plots the $NUV-r$ colour (top left),
RSF fraction (bottom left), mean metallicity (top right) and mean
[Mg/Fe] (bottom right) as a function of $B$ and $C$. We find that,
in the high $C$ regime ($C>100$), which we take to be the
traditional definition of monolithic collapse (see arguments
above), models which fit the metallicity and alpha-enhancement
expected in local early-type galaxies generally do \emph{not}
produce the blue colours of large blue early-types. Models with
$B<0.3$ reproduce the full spectrum of average metallicities and
alpha-enhancements observed in large early-type galaxies. However,
the $NUV-r$ colour predicted by these models is typically above
$\sim 6$, because RSF fractions contributed by stars within the
last Gyr (which dominate the $UV$ flux) are too low ($<0.005$),
compared with blue galaxies predicted in merger models ($\sim
0.03$). To illustrate this, we compare, in Figure
\ref{fig:compare_sam_mono}, a `monolithic' SFH ($B=0$, $C=100$) to
the \emph{average} SFH of large ($>L_*)$, blue ($NUV-r<5.5$)
early-type galaxies predicted in the merger model. The top panel
shows the non-cumulative SFH, while the bottom panel compares the
cumulative SFH. Due to the inadequate RSF fractions, blue colours
can \emph{only} be produced in a traditional monolithic scheme by
invoking high values of $B$ which leads to high metal loss and a
metal-poor galaxy. However, such metal-poor models do not fit the
observed metallicities of large early-type galaxies in the nearby
universe.

It is, of course, conceivable that the expelled gas, parametrised
by $B$, does not escape the galaxy potential well completely but
is re-accreted by the galaxy over the dynamical timescale of the
galaxy. However, recalling that stellar mass which dominates the
rest-frame $UV$ flux is created within the last Gyr of lookback
time, we find that for reasonable dynamical timescales, expelled
gas falling back into the galaxy will have a negligible effect. We
illustrate this using a simple analytical argument. Assuming that
the entire expelled fraction ($B$) falls back into the galaxy and
that this `reservoir' of initially expelled gas empties
`exponentially' we have

\begin{equation}
g(t) = B\exp({-t/\tau})
\end{equation}

where $g(t)$ is the expelled gas contained in the reservoir and
$\tau$ is the dynamical timescale of the galaxy. Therefore,
between times $t=t_1$ and $t=t_2$ the amount of gas falling into
the galaxy is given by

\begin{equation}
g(\triangle t) = B[\exp({-t_1/\tau})-\exp({-t_2/\tau})]
\end{equation}

To make a significant impact on the $UV$ colour of the galaxy,
this infalling material must contribute $\sim 3$ percent of the
stellar mass of the galaxy. Assuming that the infall of expelled
gas starts promptly after the main mass of the galaxy is in place,
i.e. $\sim 5$ Gyrs after the beginning of star formation (see
Figure \ref{fig:beq0}), the gas which produces the $UV$ emitting
stars falls in between $t=7$ and $t=8$; assuming the Universe is
$\sim 13$ Gyrs old. Hence, $UV$ emitting stars less than 1 Gyr old
will form a stellar mass fraction approximately equal to

\begin{equation}
B[\exp({-7/\tau})-\exp({-8/\tau})]
\end{equation}

which is required to be $\simeq 0.03$, to produce the blue
observed colours. However, it is difficult to achieve such mass
fractions - reasonable dynamical timescales close to what is
expected for large early-type galaxies ($\sim 1$ Gyr) provide
stellar fractions which are two orders of magnitude lower than the
required $\sim 3$ percent. The stellar fraction is maximum ($\sim
1.5$ percent) for $\tau \sim 7$ Gyrs which is unrealistic for
large early-type galaxies. Robust \emph{observational} constraints
on the amounts of recycled gas will aid the type of analysis
presented in this section. Although recent studies
\citep[e.g.][]{Young2005}, which compare the distribution of
specific angular momenta of gas and stars, do indicate that at
least some of the gas in early-type galaxies is created from
stellar mass loss, it is still unclear what fraction of the
detected gas may have an \emph{external} origin. In particular,
the gas fractions must eventually be correlated with the
intergrated $UV$ fluxes from these systems, to determine the
contribution of internally sourced gas to the $UV$ luminosity of
the early-type galaxy in question - such a study is currently in
progess using GALEX and SAURON data and results may become
available in the near future (Martin Bureau, private
communication).

However, based on the analysis presented in this section we
conclude that monolithic evolution, where RSF is driven
\emph{solely} by recycled gas from stellar mass loss, is not a
viable channel for the production of large blue early-type
galaxies.





\section{Conclusions}
We have studied $\sim$ 2100 early-type galaxies in the SDSS DR3
which have been detected by the GALEX medium depth (MIS) survey,
in the redshift range $0<z<0.11$. The early-type sample has been
selected through careful morphological inspection, with
potentially $UV$-contaminating AGN removed through the use of both
optical spectral data (from the SDSS) and radio data (from the VLA
FIRST survey). At a 95 percent confidence level, \emph{at least}
$\sim$ 30 percent of early-type galaxies in this sample have
optical and $UV$ photometry consistent with \emph{some} recent
star formation within the last Gyr.

Our analysis indicates that, while optical CMRs cannot distinguish
between early-type galaxies which have had star formation within
the last Gyr and those which havent, the $NUV$ CMR is an excellent
diagnostic of RSF - any galaxy with $NUV-r<5.5$ has a high
likelihood of containing recent star formation (RSF), even after
taking into account the possibility of a contribution to the $NUV$
spectrum from $UV$ upturn flux.


Comparison of the observations to predictions of a semi-analytical
$\Lambda$CDM hierarchical merger model yields good quantitative
agreement (across the $UV$ and optical spectrum), if we assume
that (a) very young stars ($<30$ Myrs old) are contained in birth
clouds which increase dust extinction by a factor of $\sim 3$
compared to the ISM alone and (b) that star formation is driven by
random (pseudo-instantaneous) starbursts which are Poisson
distributed in time, so that there is a small timelag between the
last starburst and the observation of the galaxy - the timelags
are distributed exponentially in time, with maximum and minimum
values of 200 and 20 Myrs respectively.

Combining our parametric analysis of the $UV$ + optical photometry
of the observed early-types, with the properties of the predicted
population in the semi-analytical model, we conclude that
early-type galaxies in the redshift range $0<z<0.11$ are likely to
have $\sim$ 1 to 3 percent of their stellar mass in stars less
than 1 Gyr old. The ($V$-band luminosity weighted age) of this
recent star formation is $\sim$ 300 to 500 Myrs.

Finally, we find that monolithically evolving galaxies, where RSF
can be produced solely from recycled gas due to stellar mass loss,
do not exhibit the blue colours ($NUV-r<5.5$) seen in some large
early-type galaxies in our observed sample. While the degeneracy
between the monolithic and merger paradigms cannot be broken for
red early-types even with $UV$ photometry, a monolithic scenario
is a very unlikely channel for the evolution of large blue
early-type systems. Such blue galaxies require \emph{additional}
fuel for star formation which must necessarily have an external
origin.


\section{Acknowledgements}
We are indebted to Ignacio Ferreras for providing the chemical
enrichment code which forms the basis of Section 6 and for
numerous stimulating discussions during the course of this study.
We warmly thank Chris Wolf for providing high-resolution COMBO-17
images which formed an integral part of our morphological
classification process. We also thank Mariangela Bernardi for her
generous help in the initial stages of this project, for providing
the DR2 versions of her SDSS early-type catalog prior to
publication, and many interesting discussions. We are grateful to
Jeremy Blaizot for his extensive help with the GALICS model and to
Andr\'es Jord\'an, Joseph Silk, Roger Davies and Andrew Benson for
many useful comments regarding this work. SK acknowledges PPARC
graduate DPhil scholarship PPA/S/S/2002/03532.



\nocite{Bernardi2003a} \nocite{Bernardi2003b}
\nocite{Bernardi2003c} \nocite{Bernardi2003d} \nocite
{Falcon-Barroso2005} \nocite{Martin2005}


\bibliographystyle{chicago}
\bibliography{references}


\end{document}